\documentclass[twocolumn,english,aps,prb]{revtex4-1}
\usepackage[T1]{fontenc}
\usepackage[latin9]{inputenc}
\usepackage{color}
\usepackage{amsmath}
\usepackage{amssymb}
\usepackage{graphicx}
\usepackage{babel}

\begin{document}

\title{Electronic properties of the FeSe/STO interface from first-principle
calculations}

\author{A.~Linscheid}
\affiliation{Department of Physics, University of Florida, Gainesville, Florida 32611, USA}

\begin{abstract}
Compared to experiment, the calculated bulk lattice constant of FeSe
is too small in LDA, PBEsol and PBE type exchange-correlation ($xc$)
functionals even though the mismatch decreases from LDA to PBE. In
bulk SrTiO$_{3}$ (STO) on the other hand LDA over-binds, PBE under-binds
and PBEsol agrees best with experiment. With the errors in opposite
directions, the strain in the FeSe monolayer on STO depends on the
$xc$ functional and, especially in the non-magnetic state, is strongly
overestimated. In this work we investigate the influence of the $xc$
functional and magnetism in density functional calculations on electronic
bonding properties and charge transfer in the FeSe monolayer on a
STO substrate. Furthermore, we consider the effect of oxygen-vacancies
within the virtual crystal approximation. In agreement with earlier
work, we find that the band structure of the checkerboard antiferromagnetic
configuration agrees best with experiment where, in addition, the
relative strain on FeSe is best described. For a small vacancy concentration,
in agreement with recent experiment, the FeSe monolayer Ti layer distance
increases while for higher concentrations it decreases. 
\end{abstract}
\maketitle

\section{Introduction}

The discovery of superconductivity (SC) in the single monolayer of
FeSe on ${\rm SrTiO}_{3}$ has triggered an enormous excitement in
the community, first because of the very high critical temperature
$T_{c}$ with an ARPES gap closing at $\sim65$K and second because
the missing hole pocket at the Brillouin zone center. The absence
of the hole pocket directly challenges the usual explanation of SC
in the Fe-based superconductors by repulsive interactions that causes
a sign changing gap function on the electron-like and the hole-like
Fermi surface sheets of these materials\cite{Wang2012,Liu2012,He2013}.
Measurements of the anisotropy of the gap function on the remaining
electron pocket have ruled out a dominant $d$-wave character \cite{Fanetal_impurities_Fe-STO15,Zhang2015}.
In the search for the origin of SC in this particular material it
is worth nothing that the electronic structure of the FeSe monolayer
is very similar to the FeSe intercalates \cite{Qian2011,Burrard-Lucas2013}
or the electron doped multi-layer surfaces \cite{Miyata2015} where
ones also finds a critical temperature of around 40-50K with no apparent
sign of attractive interactions in these system. Thus, we are left
with the challenge to explain an apparent $s$-wave gapped 40K superconductor
with repulsive interactions.

One suggested explanation, recently proposed by two groups \cite{Linscheid2016,Mishra2016}
is that $s_{\pm}$ pairing survives, even though the hole band is
$\sim60$meV below the Fermi level and thus ``incipient''. So far,
both groups have used a very simplified model of the electronic band
structure. The position of the hole band extremum is crucial since
with the supposedly strong interactions in these systems, a magnetic
transition is induced as soon as it approaches the Fermi level. In
experiment this position corresponds to the effective doping level
that can be controlled either directly by potassium dosing \cite{Song2016}
or liquid gating techniques \cite{Hanzawa2015,XHChen2015} or, effectively,
by annealing steps \cite{He2013}. The doping in the latter case is
discussed to be caused by the occurrence of oxygen vacancies in the
interface\cite{Wang2016,Shanavas2015,Chen2016}.

While the cause for SC in the FeSe intercalates is likely of purely
electronic origin, in the monolayer of FeSe on STO with a higher $T_{c}$,
on the other hand, there have been clear signs of electron-phonon
interactions. Ref.~\onlinecite{Lee2014} found replicas of the Fe
$d$-orbital bands at the Fermi level which follow the dispersion
of electronic bands with an energy offset of $100$meV, which matches
the energy of one optical oxygen modes in the STO substrate. The occurrence
of these replica bands can be explained by noting that the polar oxygen
mode will induce a dipole coupling of the electronic states in FeSe.
In the model of Ref.~\onlinecite{Lee2014}, the coupling away form
zero momentum transfer falls of exponentially with a length scale
controlled by the distance of the FeSe from the O ions in the STO
substrate making the calculation of this parameter crucial\cite{Lee2015}.

As was pointed out by Ref.~\onlinecite{Lee2014} such a small momentum
or so called forward scattering coupling is attractive in almost all
pairing channels. In line with this conclusion, Ref.~\onlinecite{Chen2015}
showed that such a low momentum transfer will induce only intra-band
coupling which will further enhance a possible incipient $s_{\pm}$
pairing state. This can explain the further enhancement of $T_{c}$
in the monolayers as compared to the purely electronic systems. A
further theoretical understanding of this mechanism and, in particular,
an extension to a more realistic electronic structure using computational
methods would be desirable.

Density functional methods\cite{HohenbergKohn1964,KohnSham1965} have
established a good starting point for realistic calculations in the
Fe-based superconductors \cite{Singh2008,Mazin2008,Kimber2009} (see
Ref.~\onlinecite{HirschfeldKorshunovMazinGapSymmetryAndStructureOfFeBasedsuperconductors2011}
for a review). However, to accurately describe the electronic structure
of most FeSe systems remains difficult. DFT calculations, in particular
within the generalized gradient approximation, have a tendency to
overestimate the magnetic moment\cite{HirschfeldKorshunovMazinGapSymmetryAndStructureOfFeBasedsuperconductors2011,Mazin2008a}.
For example, LDA\cite{Kumar2012} and PBE\cite{Ricci2013} predict
an antiferromagnetic ground state of bulk FeSe at the experimental
lattice configurations while the system is paramagnetic in experiment.
Performing a relaxation in the paramagnetic state leads to a strongly
underestimated Fe-Se bonding length and van der Waals interactions
are important to accurately predict the layer separation\cite{Ricci2013}.
Also for bulk FeSe the electron and hole pockets turn out too large
which may be explained by the underestimation of the repulsion between
the electron and hole pockets within DFT\cite{Ortenzi2009}.

In spite of these inaccuracies, several groups have calculated properties
of the monolayer FeSe on STO from first principles. Ref.~\onlinecite{Bazhirov2013}
found that for a free standing layer the checkerboard antiferromagnetic
structure agrees best with experiment. Shanavas et al.~\cite{Shanavas2015}
studied the FeSe monolayer while accounting for the STO substrate.
They consider vacancies in the system by replacing 20\% of the surface
O with F to simulate oxygen vacancies on the level of a virtual crystal
approximation (VCA). While this does induce a charge transfer to the
surface, it is found that the Fe $d$ orbitals that constitute the
hole pocket remain at the Fermi level.

Ref.~\onlinecite{Chen2016} used a super-cell approach to discuss
the effect of an oxygen vacancy on the band structure within the non-magnetic,
the checkerboard- and the stripe antiferromagnetic ordering. The super-cell
approach has been used before for to study the pure STO surface\cite{Shen2012}.
In a further step, Y. Wang et al.~\cite{Wang} computed the electron-phonon
coupling in the ${\rm Ti}-{\rm O}$ terminated slab and in a $2\times1$
super cell with an oxygen vacancy. While the focus of this work is
on the resolution of the momentum dependence of the electron-phonon
coupling, it is also found that the vacancy caues a stronger binding
between the FeSe layer and the substrate.

Moreover, S.~Coh et al.~\cite{Coh2015} and Li et al.~\cite{Li_phonon}
have calculated the electron-phonon coupling of the interface and,
recently, S.~Coh et al.~\cite{Coh2016} proposed a superstructure
of this interface to further enhance the phonon coupling.

Most calculations in this context have used the Perdew-Burgke-Ernzerhof
(PBE) density functional\cite{Perdew1996} and all studies seems to
agree that introducing vacancies binds the FeSe layer closer to the
substrate and electron-dopes the FeSe layer.

In a recent experimental study \cite{Li2015}, it was discovered that
upon annealing where the system turns from non-superconducting to
superconducting, the binding distance of the FeSe layer, in fact,
increases. This finding is in contradiction to the theoretical calculations
even though comparison is difficult because the actual vacancy content
in experiment is not easily determined. Ref.~\onlinecite{Li2015}
suggests that the interface is, in fact, FeSe on STO terminated by
a double layer of ${\rm Ti}-{\rm O}$. Moreover, the monolayer is
shown to cover the STO substrate incommensurately. 

It is known that PBE tends to overestimate the binding distance while
LDA type functionals are known to underestimate it \cite{Perdew2008}.
This insight seems particularly important because it is common practice
to fix the $x-y$ lattice constant to experiment while one implicitly
allows the $z$ direction relaxation of the slab since the cell contains
a large part of vacuum. This may cause a very large non-uniaxial pressure
in the calculations.

In this work, we revisit the bulk calculations for both STO and FeSe
and point out that the dismatches of lattice constants in $x-y$ go
into opposite directions - predicted lattice parameters within the
layer are too small in FeSe while in STO they obey the usual trend.
As we find that PBEsol\cite{Perdew2008} is best for STO and PBE in
an antiferromagnetic state is best for FeSe we compute the interface
in both functionals. As oxygen vacancies play a key role in this system,
we further investigate the structure and the charge transfer as a
function of vacancy content within the VCA which allows us to treat
small vacancy concentrations. We find for low concentrations that
the layer moves away from the substrate and the doping effect on Fe
is small while for higher concentrations the bond length Se-Ti shortens
and charge is transferred to the FeSe layer.

\section{Computational details}

We model the surface by placing an FeSe monolayer on a slab of 3 layers
of ${\rm SrTiO}_{3}$. Periodic replicas in $z$ direction are separated
by $>10${\AA} of vacuum to eliminate axillary interaction. All subsequent
DFT calculations were performed with Quantum ESPRESSO \cite{QE-2009}
using ultra-soft GBRV pseudo potentials \cite{Garrity2014}. Accurate
convergence was achieved within a plane wave cutoff of 50Ry and 200Ry
for wave function and charge density unless otherwise noted. We sample
the Brillouin zone using a $6\times6\times1$ Monkhurst-Pack grid
\cite{Monkhorst1976} with a Methfessel-Paxton smearing \cite{Methfessel1989}
of 0.02Ry. We use a very strict condition of $5\times10^{-5}{\rm Ry}/{\rm au}$
and $10^{-6}$Ry for force and total energy convergence during relaxations.
Especially when oxygen vacancies are considered, the asymmetric slab
shows a dipole moment. In order to remove spurious interaction with
the periodic replicas, an axillary compensating dipole is introduced
in the vacuum. Ref.~\onlinecite{Zhou2016} concluded that the the
octahedral, ferro-electric distortion is suppressed near the surface
and in general not important for the electronic structure. Based on
this insight, we approximate the ferroelectric distorted STO with
a cubic unit cell.

\section{Bulk calculations}

\begin{table}
\begin{centering}
\begin{tabular}{|c|c|c|c|c|c|c|c|}
\hline 
 & \multicolumn{2}{c|}{LDA} & \multicolumn{2}{c|}{PBEsol} & \multicolumn{2}{c|}{PBE} & Exp\\
\hline 
 & NM & AFM & NM & AFM & NM & AFM & \\
\hline 
$a_{{\rm {\scriptscriptstyle STO}}}${[}\r{A}{]} & 3.842 &  & 3.892 &  & 3.937 &  & 3.907\cite{Cao2000} \\
\hline 
$a_{{\rm {\scriptscriptstyle FeSe}}}${[}\r{A}{]} & 3.589 & - & 3.623 & 3.631 & 3.685 & 3.716 & 3.765\cite{Louca2010}\\
\hline 
$b_{{\rm {\scriptscriptstyle FeSe}}}${[}\r{A}{]} & 3.589 & - & 3.623 & 3.623 & 3.685 & 3.709 & 3.754\cite{Louca2010}\\
\hline 
$c_{{\rm {\scriptscriptstyle FeSe}}}${[}\r{A}{]} & 5.301 & - & 5.476 & 5.544 & 6.336 & 6.206 & 5.479\cite{Louca2010}\\
\hline 
$h_{{\rm {\scriptscriptstyle FeSe}}}${[}\r{A}{]} & 1.364 & - & 1.370 & 1.397 & 1.377 & 1.447 & 1.462\cite{Louca2010}\\
\hline 
$\vert m\vert/{\rm Fe}\,[\mu_{{\rm {\scriptscriptstyle B}}}]$  &  & 0 &  & 1.669 &  & 2.362 & \\
\hline 
\end{tabular}
\par\end{centering}

\caption{Structural parameters for the bulk materials cubic STO and FeSe for
the three functional in comparison with experiment. For FeSe we allow
a checkerboard antiferromagnetic order (AFM). Within LDA, the structure
converged to one that has a non-magnetic (NM) ground state.\label{tab:Structbulk}}
\end{table}
In preparation for the calculation of the FeSe monolayer on STO interface,
we consider the individual bulk materials in this section. The cutoff
for FeSe could be reduced to 45Ry and 180Ry for wavefunction and charge
density with a $k$ point sampling of $6\times6\times4$ and $4\times4\times4$
for STO. We performed a relaxation for all three functionals while
starting from a tetragonal phase in the non-magnetic (NM) phase or
an orthorhombic and antiferromagnetically broken symmetry. The resulting
structures are given in Tab.~\ref{tab:Structbulk}. The relaxation
within the LDA always lead to the non-magnetic structure, i.e. the
broken symmetries were restored to numerical accuracy. 

Results for STO are more in line with the usual hierarchy of bonding
lengths where LDA underestimates, PBE overestimates and PBEsol is
closest to experiment. In FeSe, however, all functionals underestimate
the in-plane lattice constant, especially if the NM state is considered.
Calculated lattice constants for LDA are in good agreement with earlier
work \cite{Winiarski2012}. Ref.~\onlinecite{Ricci2013} finds that
the stripe antiferromagnetic state, that has in fact the lowest energy,
shows an even better agreement of the $a_{{\rm {\scriptscriptstyle FeSe}}}$
and $b_{{\rm {\scriptscriptstyle FeSe}}}$ lattice constants. However,
since earlier work on the FeSe/STO interface agrees that the electronic
structure of the checkerboard antiferromagnetic state agrees best
with experiment, we focus on this phase and its bulk properties as
calculated with DFT methods. In the following, we shall always refer
to the checkerboard order as the AFM state. Ref.~\onlinecite{Ricci2013}
also points out that van der Waals interactions are crucial to compute
out of plane, $c_{{\rm {\scriptscriptstyle FeSe}}}$, lattice constant
correctly. That the same conclusion holds for the monolayer on STO
is not immediately clear, since, especially if there is charge transfer,
other effects are likely dominant.

From this analysis, we conclude that we can expect the structure of
the STO to be best described within the PBEsol functional while PBE
will perform better for the FeSe monolayer. In the following we will
compare the performance of these functionals within the AFM and NM
state while we include oxygen vacancies at the last stage.
\begin{figure}
\begin{centering}
\includegraphics[width=0.75\columnwidth]{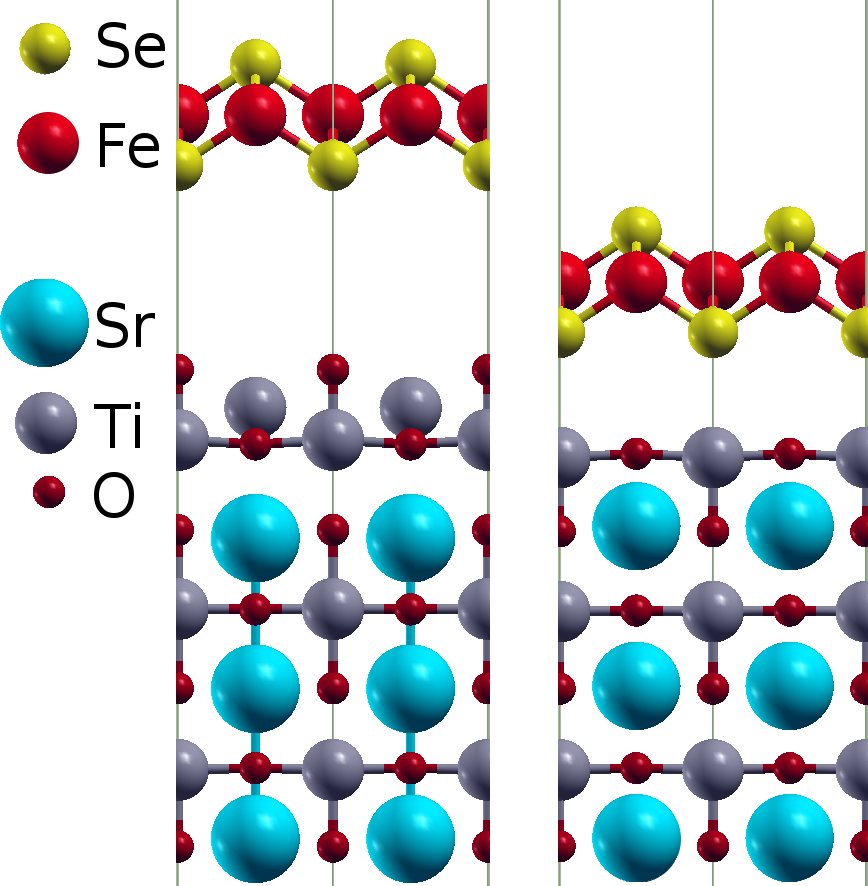}
\par\end{centering}

\caption{Relaxed structure of the double ${\rm Ti}-{\rm O}$ layer within PBE
on the left and with the single layer ${\rm Ti}-{\rm O}$ terminated
slab on the right. Parts of the cell on the top that contain vacuum
are removed from the plot.\label{fig:DoubleTiOxStruct}}
\end{figure}

\section{Double ${\rm Ti}-{\rm O}_{x}$ layer}

In this section, we follow the suggestion of Ref.~\onlinecite{Li2015}
and consider a slab terminated by a double layer of ${\rm Ti}-{\rm O}$
below the FeSe monolayer. We model the system with a slab of 3 layers
of cubic STO and add an additional ${\rm Ti}-{\rm O}$ layer below
the FeSe. The relaxed structure within PBE is shown in Fig.~\ref{fig:DoubleTiOxStruct}.
It turns out, that the FeSe layer is poorly bound and >5\r{A} far
away from the substrate in this configuration, both in PBEsol and
PBE. In fact, it was not possible to relax the system to the required
low forces and the constrained on forces had to be reduced to $1.5\times10^{-3}$
Ry/au indicating that the minimum in the potential energy surface
is very shallow. While the binding may become closer if van der Waals
interactions are considered, we also observe a reconstruction of the
double ${\rm Ti}-{\rm O}$ where the Ti of the topmost layer moves
towards the substrate which is different from the layer found in Ref.~\onlinecite{Li2015}
where the layers keep their general structure and the separation of
the ${\rm Ti}-{\rm O}_{x}$ layers, instead, increases as compared
to bulk STO. The effect of close binding of the double ${\rm Ti}-{\rm O}$
layers is consistent within all tests performed on this system and
in disagreement with the structure reported in Ref.~\onlinecite{Li2015}.
We determine from the DFT perspective that the single ${\rm Ti}-{\rm O}_{x}$
terminated STO is the more likely configuration.

\section{Binding distance and $x-y$ relaxation}

\begin{figure}
\begin{centering}
\includegraphics[width=0.7\columnwidth]{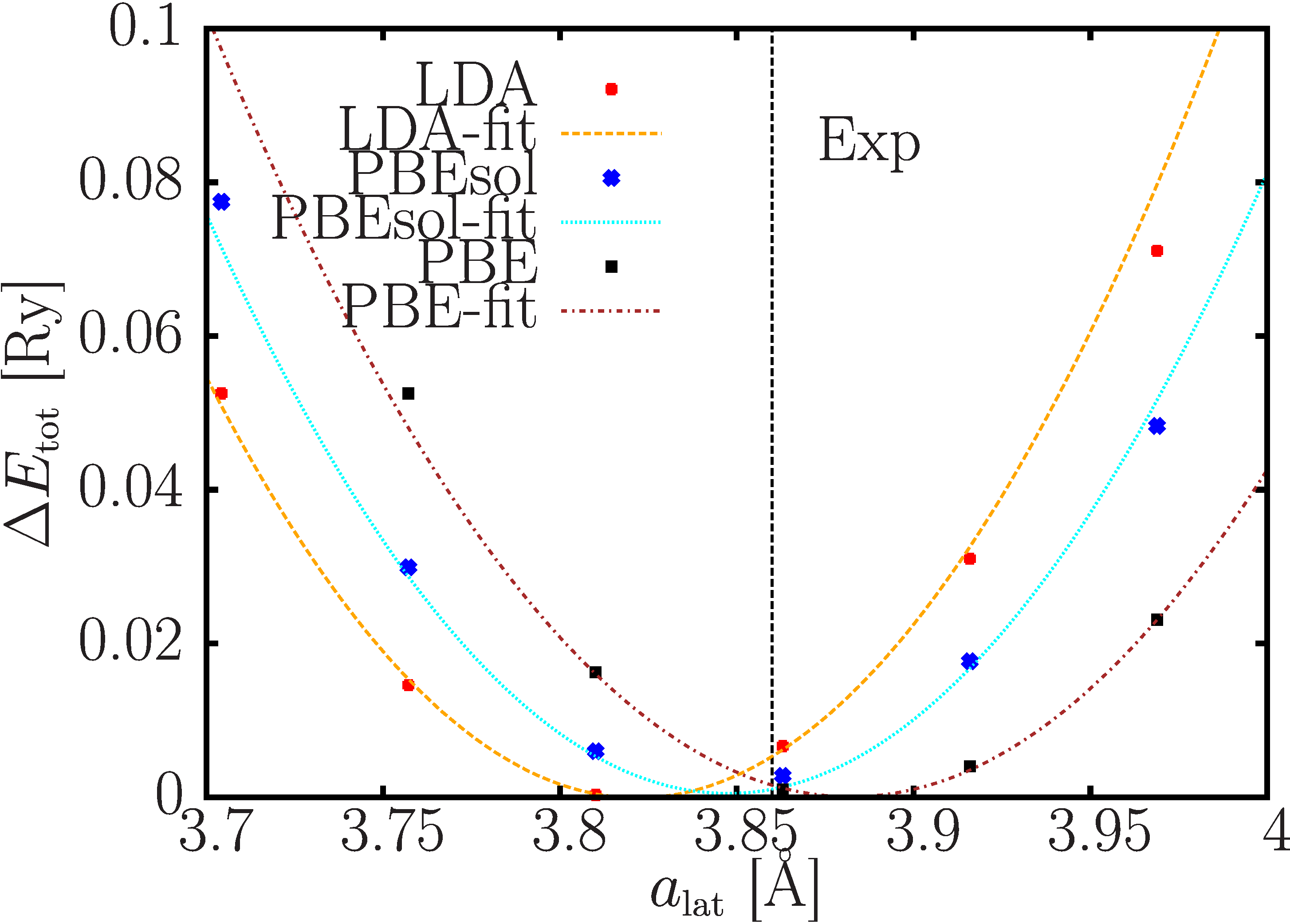}
\par\end{centering}

\caption{Total energy difference of the slab as a function of $x-y$ compression
for the $xc-$functionals LDA, PBEsol and PBE. The total energy depends
quadratically (fit) on the lattice constant and reaches its minimum
at $a_{{\rm {\scriptscriptstyle lat}}}=3.822,3.847$ and $3.880$\r{A}
for LDA,PBEsol and PBE, respectively. The experimental monolayer FeSe
lattice constant of $3.86$\r{A} is shown in the plot. \label{fig:TotalEneXYCompress}}
\end{figure}
DFT functionals usually disagree in the predicted lattice constant
with experiment up to within a few percent. In a bulk geometry, one
can let the unit cell relax to represent the minimum in total energy
which yields the equilibrium structure. In the present surface geometry,
it is particularly important to verify that a strain imposed by the
choice of unit cell does not alter the chemical bonding structure
since, as opposed to a bulk geometry, the strain on the slab in the
z direction is zero due to the vacuum. If all the chemical bonds in
the system obey the trend, choosing the experimental lattice constant
of cubic STO and using the LDA $xc$ functional will tend to stretch
the slab and planes will move closer together. In Fig.~\ref{fig:TotalEneXYCompress}
we show the total energy of the surface plus slab geometry (compare
Fig.~\ref{fig:DoubleTiOxStruct}, right) as a function of $x-y$
lattice parameter. The total energy depends quadratically on the distance
to the optimal value over a rather large range. From the quadratic
fits, we determine that equilibrium lattice constants of $a_{{\rm {\scriptscriptstyle lat}}}=3.822,3.847$
and $3.880$\r{A} for LDA,PBEsol and PBE, respectively. It is interesting
to note that bulk STO has a lattice constant of $3.9065$\r{A}\cite{Cao2000}
while Ref.~\onlinecite{Li2015} found a significantly smaller value
for the FeSe layer of $\sim3.86$\r{A}. While this implies a significant
expansion as compared to bulk FeSe (compare Tab.~\ref{tab:Structbulk}),
it is still much smaller than the bulk STO lattice constant in the
cubic phase. The reason is for this discrepancy is, in fact, that
the underlying STO is incommensurate with the surface layer and has
a slightly larger lattice constant. 

As noted earlier, all functionals considered in this work tend to
shrink the FeSe in the $x-y$ direction. The energy optimal configuration
represents a well defined compromise between these two components
of the system. In fact, reducing the STO layer number in the slab
to one, we find further reduced optimal $a_{{\rm {\scriptscriptstyle lat}}}$
values of 3.788, 3.810 and 3.841\r{A} which reflects the growing
influence of the FeSe layer.

We take the energy optimal lattice constant as a compromise between
bulk and surface stress. In Tab.~\ref{tab:NMRelax}
\begin{table}
\begin{centering}
\begin{tabular}{|c|c|c|c|c|c|c|}
\hline 
 & $h_{{\rm {\scriptscriptstyle FeSe}}}$ & $\alpha$ & $d_{{\rm {\scriptscriptstyle TiSe}}}$ & $\rho_{{\rm {\scriptscriptstyle Fe}}}$ & $\rho_{{\rm {\scriptscriptstyle Se-O}}}$ & $\rho_{{\rm {\scriptscriptstyle Se-v}}}$\\
 & {[}\r{A}{]} & {[}Deg{]} & {[}\r{A}{]} & {[}$e$/u.c{]} & {[}$e$/u.c{]} & {[}$e$/u.c{]}\\
\hline 
\hline 
LDA & 1.201 & 116.79 & 2.860 & 0.407 & -0.395 & -0.363\\
\hline 
LDA $E-o$ & 1.242 & 113.97 & 2.860 & 0.371 & -0.356 & -0.329\\
\hline 
PBEsol & 1.229 & 115.61 & 2.964 & 0.431 & -0.418 & -0.392\\
\hline 
PBEsol $E-o$ & 1.257 & 113.68 & 2.963 & 0.420 & -0.421 & -0.366\\
\hline 
PBE & 1.280 & 113.18 & 3.162 & 0.460 & -0.465 & -0.412\\
\hline 
PBE $E-o$ & 1.280 & 113.18 & 3.164 & 0.461 & -0.472 & -0.409\\
\hline 
\end{tabular}
\par\end{centering}

\caption{Comparison of relaxed NM structure and excess charge within the three
functionals at experimental and energy optimal ($E-o$) in plane lattice
constant. $\rho_{{\rm {\scriptscriptstyle Fe}}}$, $\rho_{{\rm {\scriptscriptstyle Se-O}}}$
and $\rho_{{\rm {\scriptscriptstyle Se-v}}}$ are the excess charge
according to a Bader surface analysis at a Fe site, a Se site in direction
of the substrate and a Se site in direction of the vacuum, respectively.
$\alpha$ is the Fe-Se bonding angle and $d_{{\rm {\scriptscriptstyle TiSe}}}$
the bonding distance of the Ti to Se.\label{tab:NMRelax}}
\end{table}
 we show the bonding structure and charge transfer according to a
Bader analysis. We find that the bonding distance of the FeSe layer
to the slab $d_{{\rm {\scriptscriptstyle TiSe}}}$ does not depend
on the $x-y$ stress, but significantly on the functional. With the
bonding distance being $~2.86$\r{A} and the Fe-Se distance $1.2$\r{A}
we find a step height of 5.26\r{A} in LDA that slightly increases
to 5.34\r{A} if the energy optimal $x-y$ lattice constant is used.
Both of these values are much too short as compared to the $5.5$\r{A}
measured in the original STM study \cite{Wang2012}. Comparison of
this step height within PBE and especially PBEsol is much better.
\begin{figure}
\begin{minipage}[t]{1\columnwidth}%
\begin{center}
\includegraphics[width=1\columnwidth]{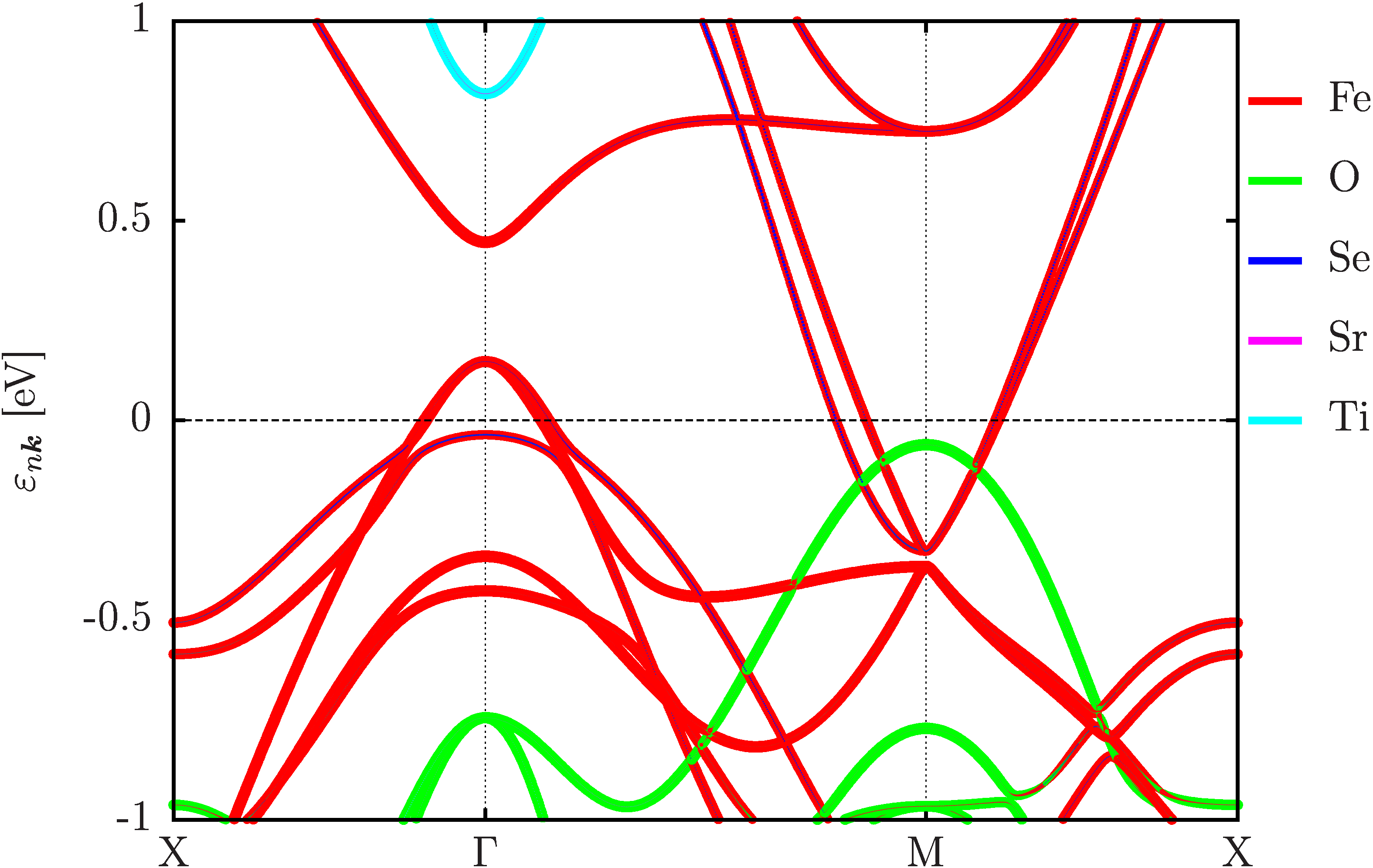}
\par\end{center}

\begin{center}
(a)
\par\end{center}%
\end{minipage}\\
\begin{minipage}[t]{1\columnwidth}%
\begin{center}
\includegraphics[width=1\columnwidth]{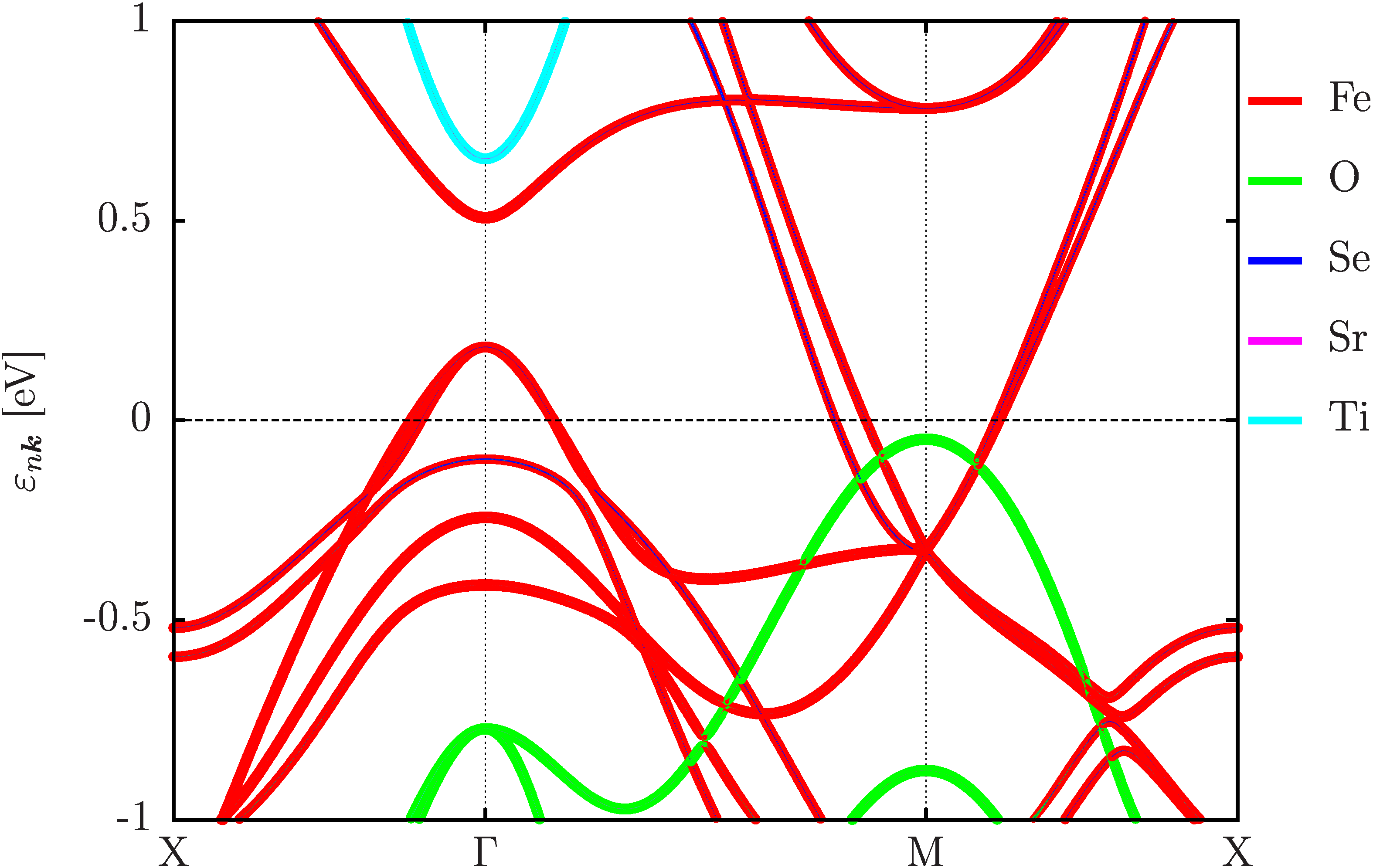}
\par\end{center}

\begin{center}
(b)
\par\end{center}%
\end{minipage}

\caption{LDA band structure at the STO bulk experimental (a) and energy optimal
(b) $x-y$ lattice constant. Colors indicate the projection on atomic
orbitals as indicated on the right of each plot and the size of each
dot corresponds to the overlap of the particular wavefunction with
the given atomic orbital.\label{fig:LDAStrain}}
\end{figure}

\begin{figure*}[t]
\begin{minipage}[t]{0.5\textwidth}%
\begin{center}
\includegraphics[width=1\columnwidth]{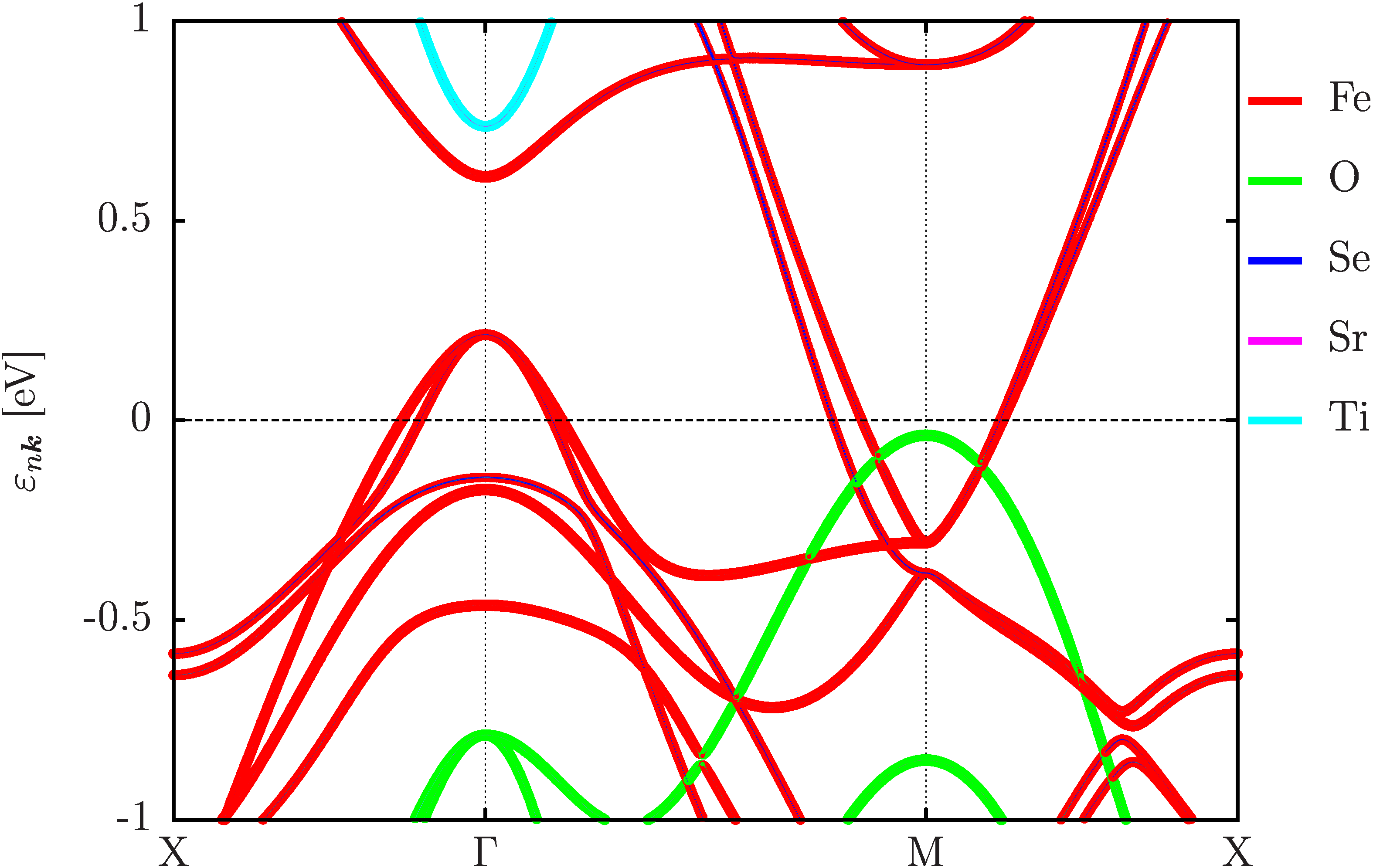}\\
a)
\par\end{center}%
\end{minipage}\nolinebreak%
\begin{minipage}[t]{0.5\textwidth}%
\begin{center}
\includegraphics[width=1\columnwidth]{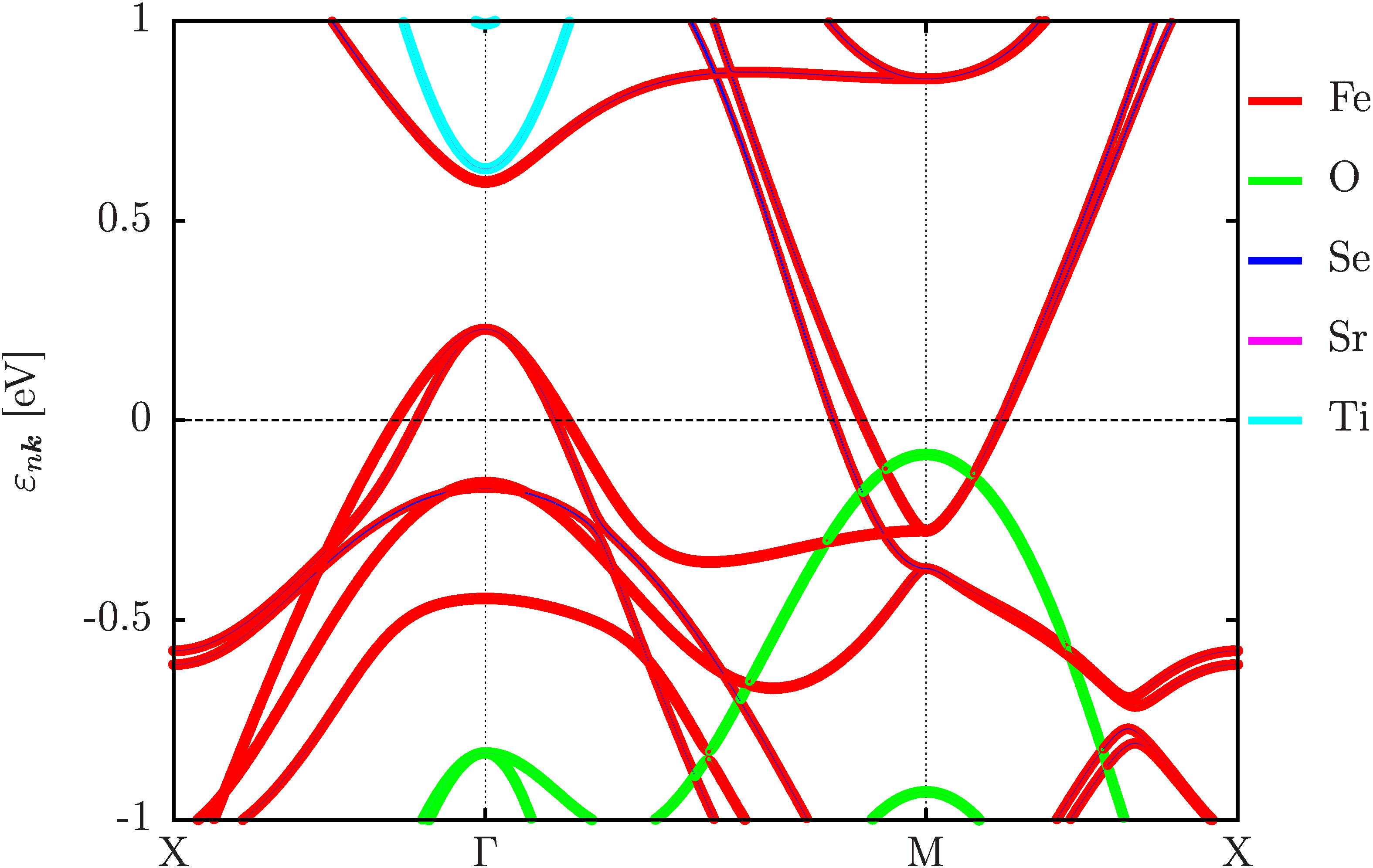}\\
b)
\par\end{center}%
\end{minipage}\\
\begin{minipage}[t]{0.5\textwidth}%
\begin{center}
\includegraphics[width=1\columnwidth]{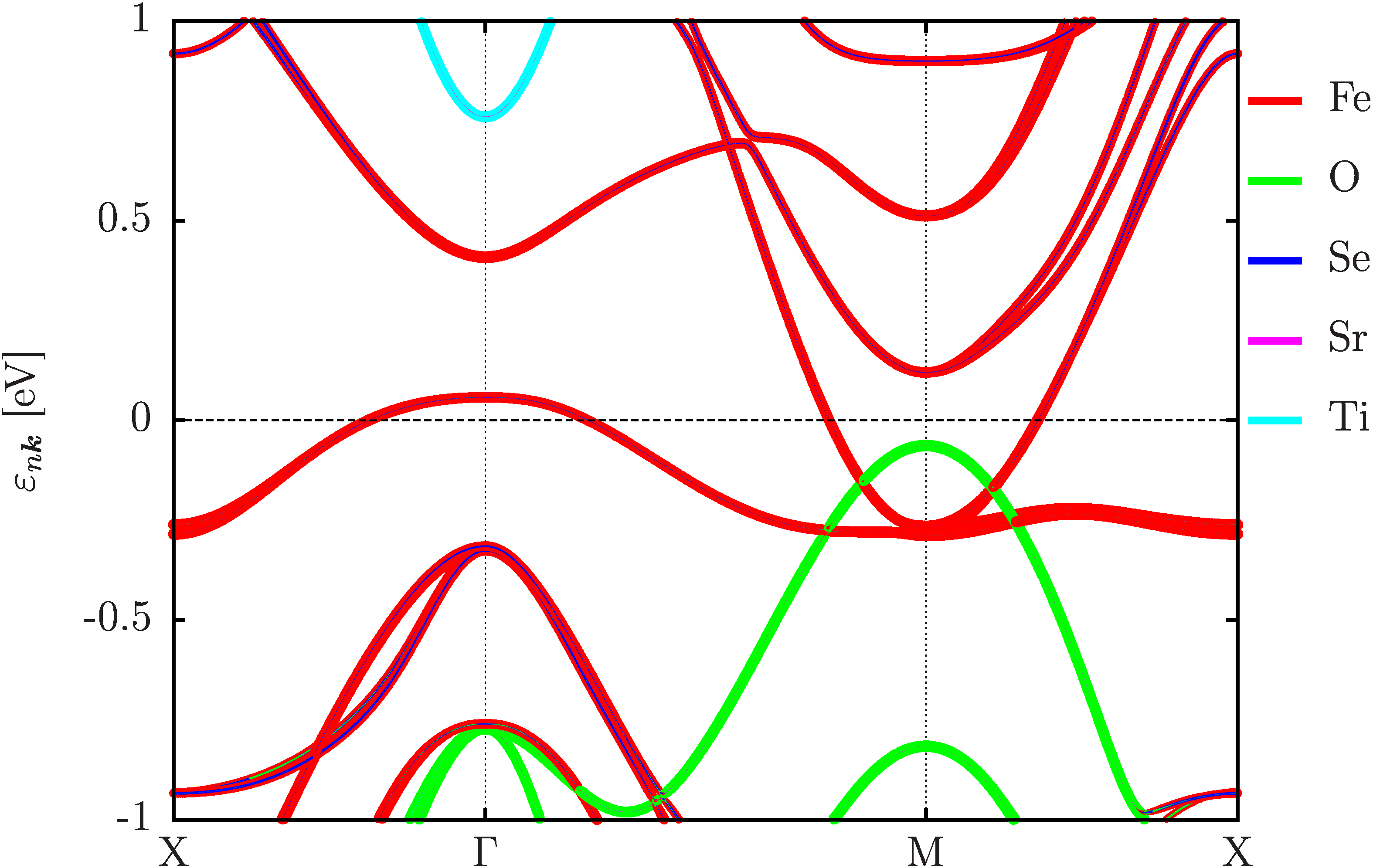}\\
c)
\par\end{center}%
\end{minipage}\nolinebreak%
\begin{minipage}[t]{0.5\textwidth}%
\begin{center}
\includegraphics[width=1\columnwidth]{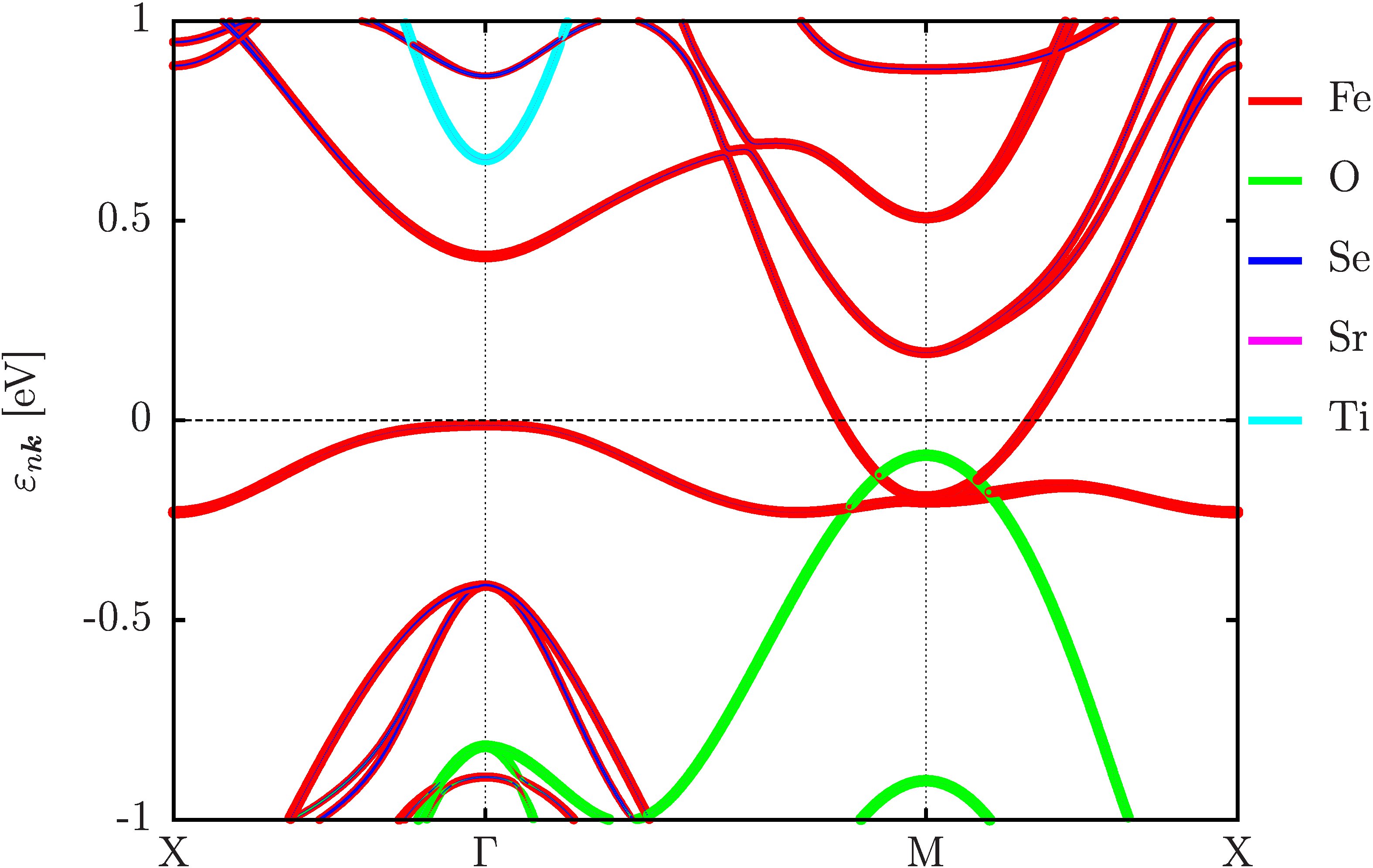}\\
d)
\par\end{center}%
\end{minipage}

\caption{Comparison of the band structure of the NM in a) and c) with the AFM
state in b) and d) within PBEsol {[}a),c){]} and PBE {[}b),c){]}.
We use the same color coding as in Fig.~\ref{fig:LDAStrain}. Lattice
parameters are energy optimal for the NM and the experimental $a_{{\rm {\scriptscriptstyle lat}}}=3.905$\r{A}
for the AFM state.\label{fig:StocPBEPBEsol}}
\end{figure*}
While the separation distance of FeSe from the STO substrate does
not depend on strain, the FeSe internal structure parameters show
significant signs of strain within PBEsol and, even more so, in the
LDA when the experimental lattice parameter are used. This is apparent
in an opening of the bonding angle $\alpha$ and a reduction of the
Se height $h_{{\rm {\scriptscriptstyle FeSe}}}$. The energy optimal
and experimental lattice parameter are very close for PBE. Performing
the calculation with the energy optimal lattice constant, we find
more similar structure parameters among the functionals. If the mismatch
between calculated layer - and substrate lattice parameter were independent
on the functional, the tensile strain should be unaffected whether
we perform the calculation in LDA, PBEsol or PBE at an energy optimal
lattice parameter. However, we still observe a tendency to open the
bonding angle and increase the Se position as compared to the Fe layer
changing the functional from PBE to PBEsol and LDA, indicating that
the relative tensile strain increases. Without O-vacancies the FeSe
layer is nearly neutral within the Bader analysis in all these calculations. 

In Fig.~\ref{fig:LDAStrain} we compare the band structure with a
projection on atomic orbitals for LDA with (a) and without (b) strain
on the unit cell. As can be seen, the hole pocket at $\Gamma$ becomes
slightly bigger as we relax the system while the heavier, filled hole
band below the Fermi level moves down in energy. At the M point, the
electron pocket slightly moves up in energy as we relax. At the example
of LDA, where the differences between energy optimal and experimental
lattice parameters are largest, we find that additional strain seems
to improve comparison with experiment although this has likely to
be attributed to smaller hoppings within the Fe layer and not a better
description of the strong interactions in this material.

Because the LDA structure describes neither the STO nor the FeSe correctly,
in the following, we focus on the two other functionals PBE and PBEsol.
Turning our attention to the magnetic state, in Tab.~\ref{tab:CAFMNoVacan}
\begin{table}[b]
\begin{centering}
\begin{tabular}{|c|c|c|c|}
\hline 
 & $h_{{\rm {\scriptscriptstyle FeSe}}}${[}\r{A}{]} & $\alpha${[}Deg{]} & $d_{{\rm {\scriptscriptstyle TiSe}}}${[}\r{A}{]}\\
\hline 
PBE & 1.371 & 109.86 & 3.129\\
\hline 
PBEsol & 1.306 & 112.45 & 2.947\\
\hline 
Exp NS\cite{Li2015} & 1.33$\pm$0.02 & 111.0$\pm$0.09 & 3.34$\pm$0.05\\
\hline 
\multicolumn{1}{c}{} & \multicolumn{1}{c}{} & \multicolumn{1}{c}{} & \multicolumn{1}{c}{}\\
\hline 
 & $\rho_{{\rm {\scriptscriptstyle Fe}}}${[}$e$/u.c{]} & $\rho_{{\rm {\scriptscriptstyle Se-O}}}${[}$e$/u.c{]} & $\rho_{{\rm {\scriptscriptstyle Se-v}}}${[}$e$/u.c{]}\\
\hline 
PBE & 0.570 & -0.560 & -0.530\\
\hline 
PBEsol & 0.527 & -0.526 & -0.483\\
\hline 
\end{tabular}
\par\end{centering}

\caption{Structure and charge in the AFM state. We use the experimental lattice
parameter $a_{{\rm {\scriptscriptstyle lat}}}=3.905$\r{A} and relax
the system for fixed unit cell. \label{tab:CAFMNoVacan}}
\end{table}
 we give the structural and charge analysis. We fix the lattice constant
to the experimental cubic STO bulk value of $3.905$\r{A}. The total
size of the 1 unit cell thick layer is computed to be 5.87 and 5.56\r{A}
which, especially for PBEsol compares well with the original measurement
of 5.5\r{A} in Ref.~\onlinecite{Wang2012}. However, comparison
with recent experimental data\cite{Li2015} reveals that the binding
distance of the layer from the substrate is too short while the structural
parameters for the layer itself appear to be somewhat in between the
results for PBEsol and PBE. Comparison may be complicated, because
for example the vacancy concentration in the normal state is unknown
while our calculation assumes the stoichiometric structure.

In Fig.~\ref{fig:StocPBEPBEsol}, we compare the magnetic and non-magnetic
electronic bands within PBEsol and PBE at their respective energy
optimal lattice configuration. We find that the differences between
PBE and PBEsol are minor. Allowing for an AFM state where we use the
experimental lattice parameters, on the other hand strongly shifts
down the hole band and the resulting band structure compare much better
with ARPES data\cite{Lee2014} than the NM data. PBE shows a slightly
higher Fermi level than PBEsol, thus making the agreement even better.
\begin{figure}
\begin{raggedleft}
\begin{minipage}[t]{1\columnwidth}%
\begin{center}
\includegraphics[width=0.7\columnwidth]{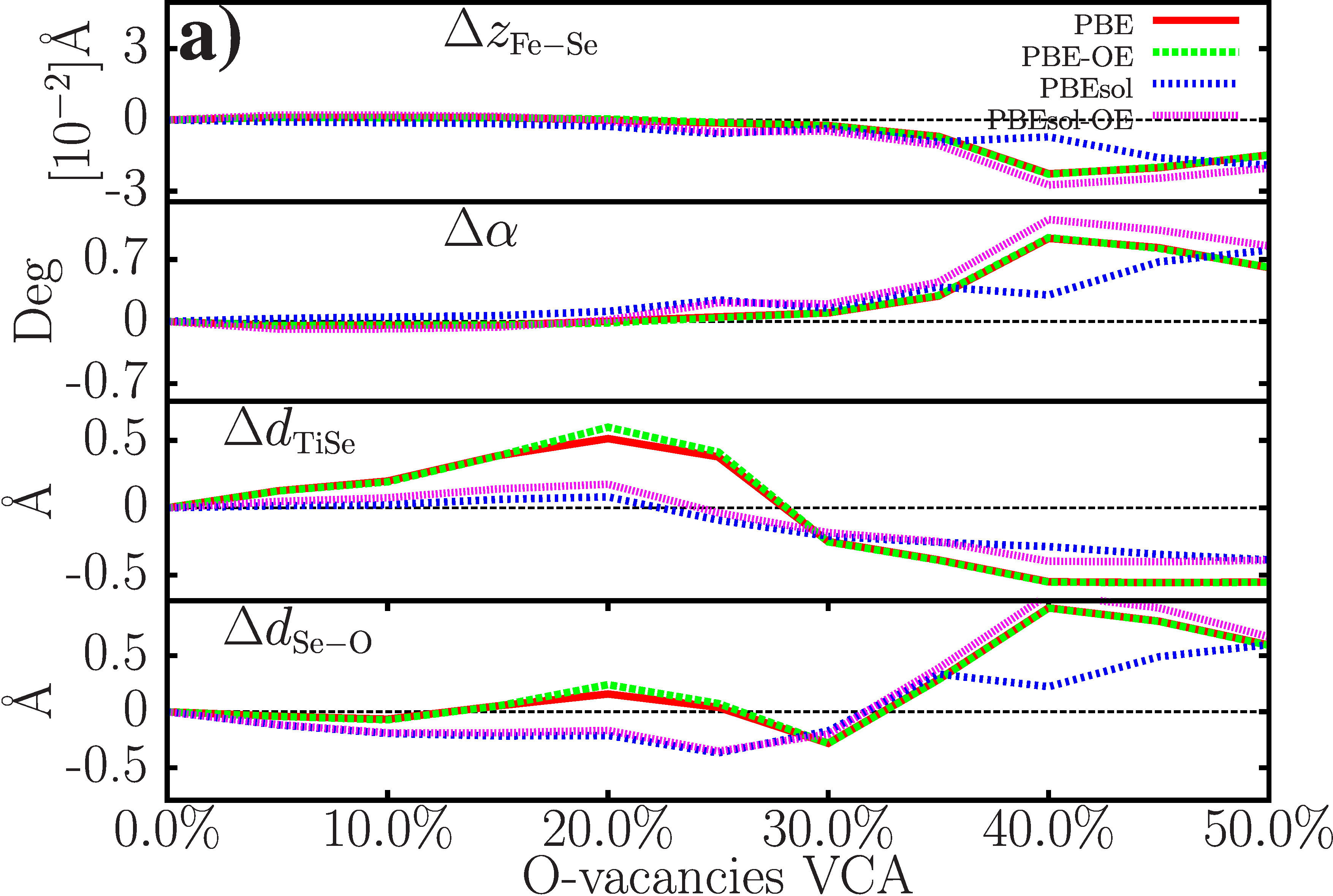}
\par\end{center}%
\end{minipage}\\
\begin{minipage}[t]{1\columnwidth}%
\begin{center}
\includegraphics[width=0.7\columnwidth]{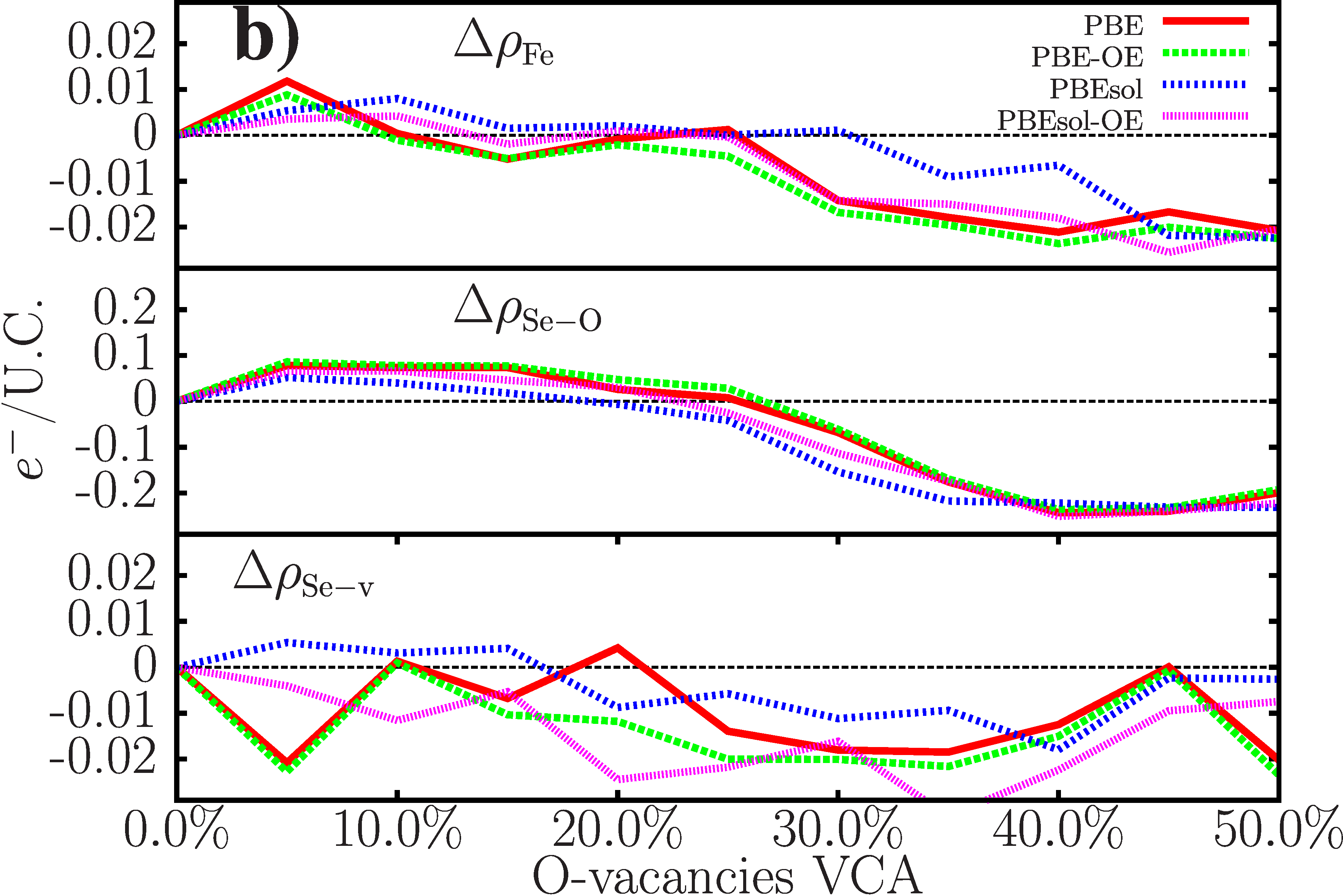}
\par\end{center}%
\end{minipage}
\par\end{raggedleft}

\caption{a) Structure data as a function of oxygen vacancies within the VCA.
b) Charge polarization associated to Fe and Se atoms are according
to a Bader analysis. We define the difference $\Delta\alpha$ as $\alpha(x\%)-\alpha(0\%)$
and for other parameters accordingly.\label{fig:StructChargeNM}}
\end{figure}
\begin{figure}
\begin{centering}
\begin{minipage}[t]{0.5\columnwidth}%
\begin{center}
\includegraphics[width=1\columnwidth]{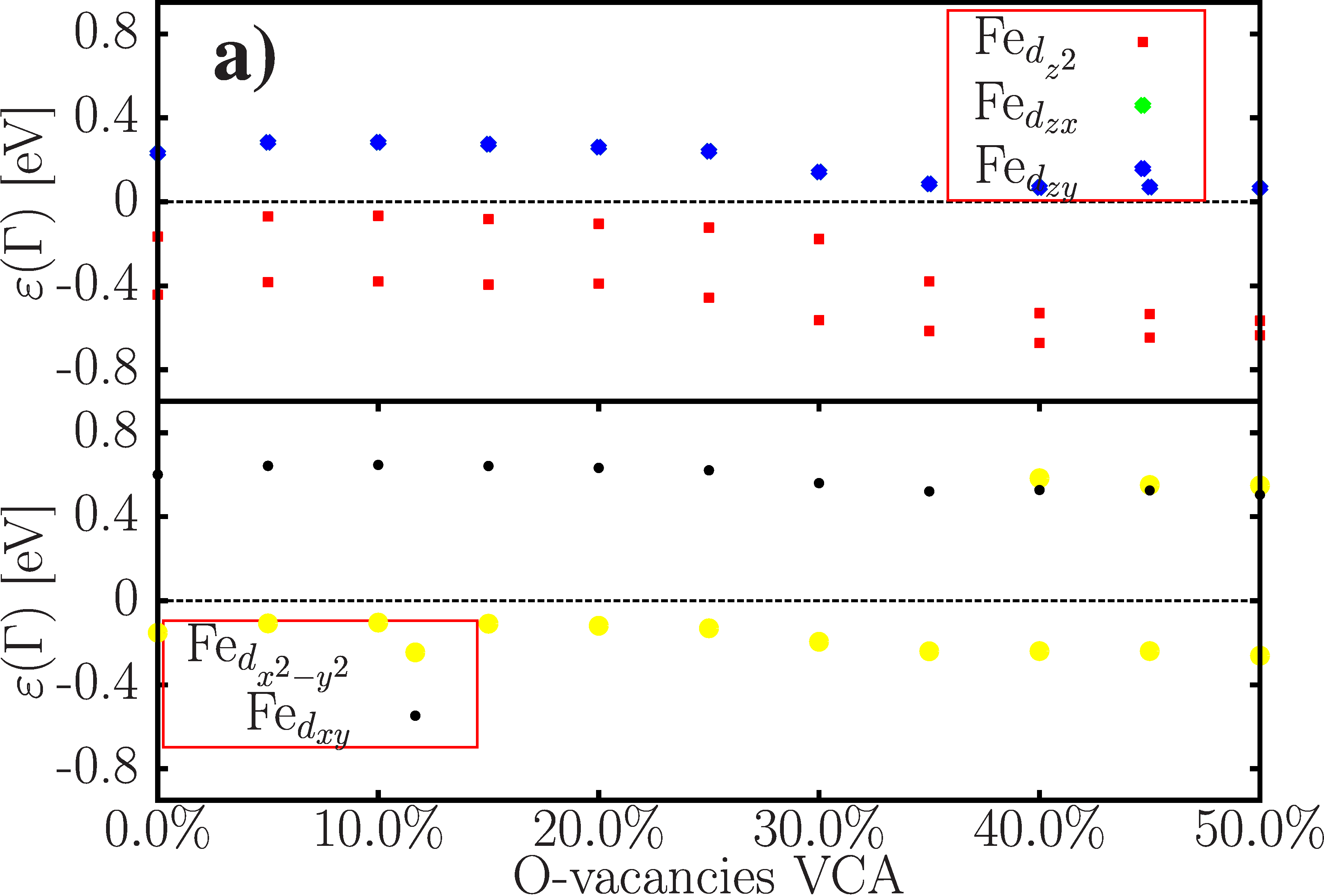}
\par\end{center}%
\end{minipage}\nolinebreak%
\begin{minipage}[t]{0.5\columnwidth}%
\begin{center}
\includegraphics[width=1\columnwidth]{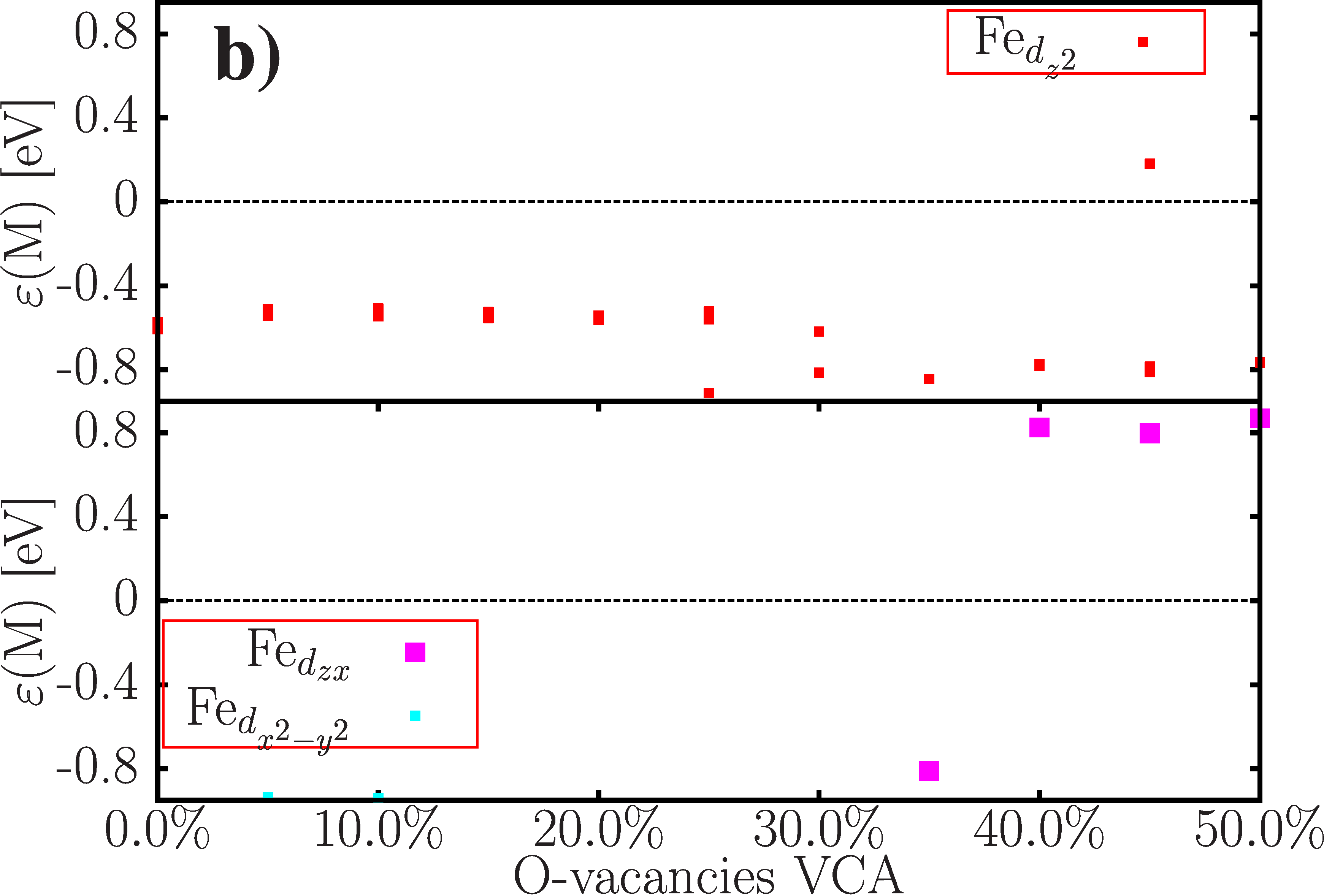}
\par\end{center}%
\end{minipage}\\
\begin{minipage}[t]{0.5\columnwidth}%
\begin{center}
\includegraphics[width=1\columnwidth]{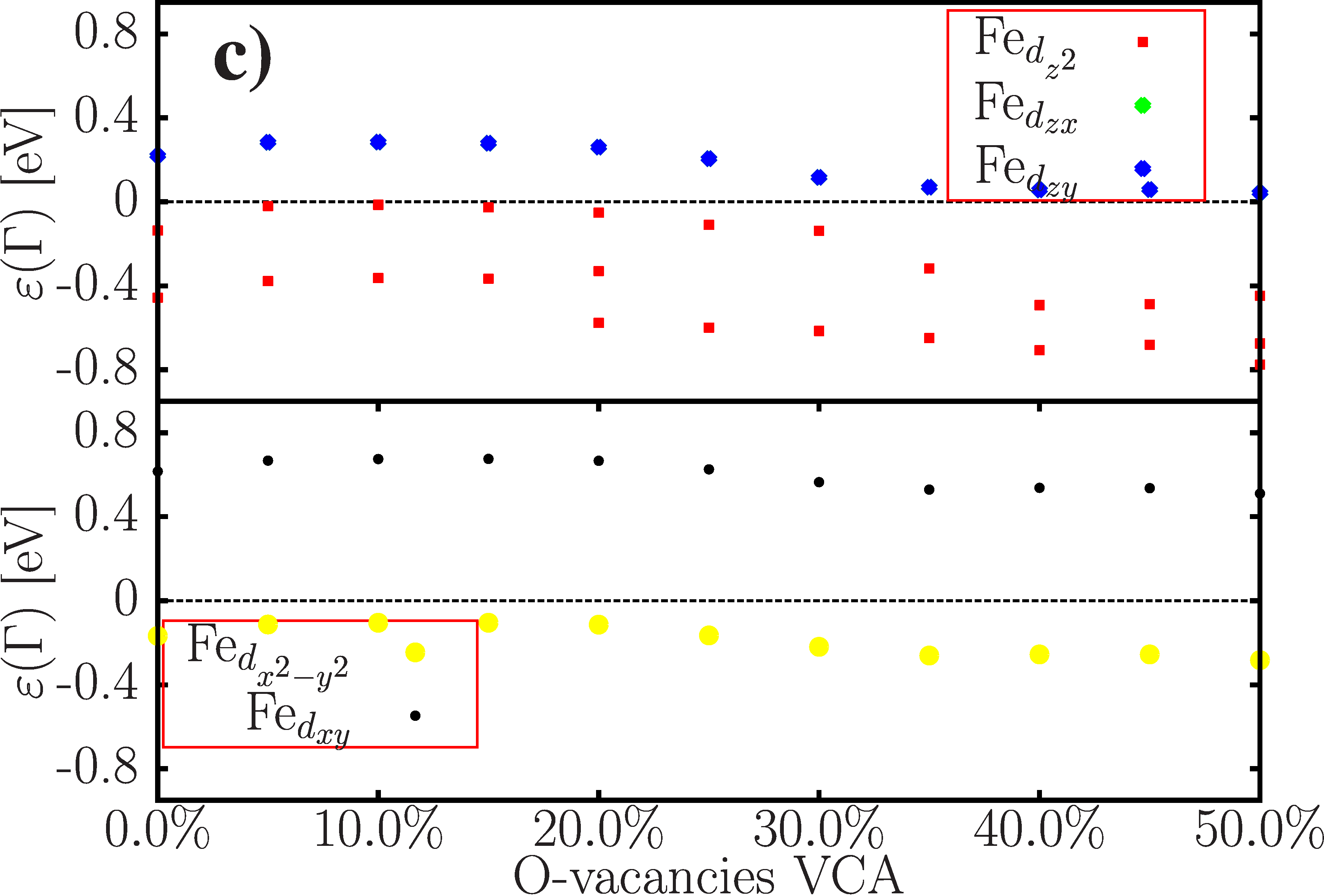}
\par\end{center}%
\end{minipage}\nolinebreak%
\begin{minipage}[t]{0.5\columnwidth}%
\begin{center}
\includegraphics[width=1\columnwidth]{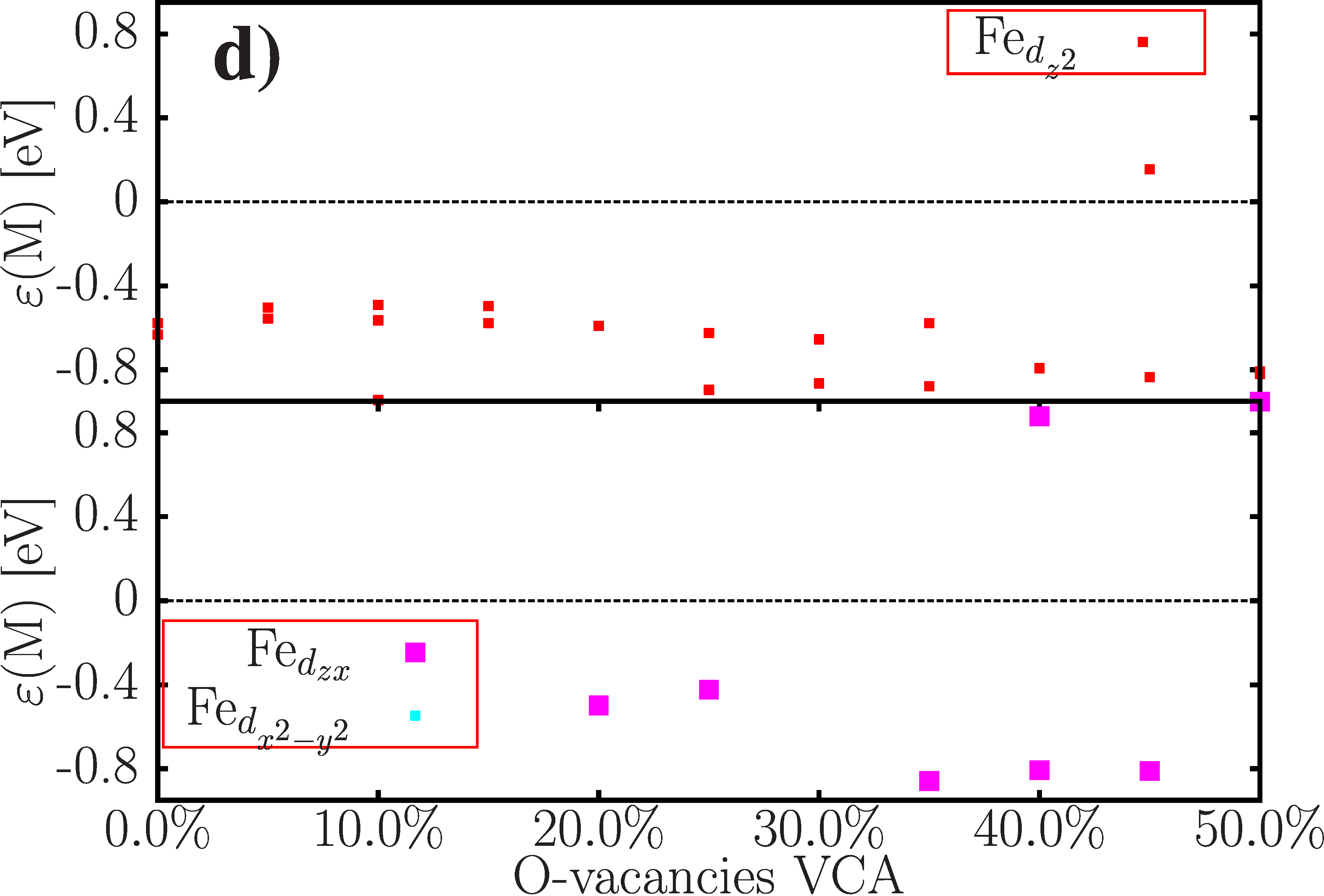}
\par\end{center}%
\end{minipage}
\par\end{centering}

\caption{Evolution of bands with a majority character of Fe at $\Gamma$ {[}a)
and c){]} and M {[}b) and d){]} point in the Brillouin zone with O
vacancy contribution. a) and b) are within PBE while c) and d) are
within PBEsol at their energy-optimal lattice parameter. Note that
we use the 2-Fe unit cell for the quantization axis of the $d$-orbitals.\label{fig:BandsNonMagn}}
\end{figure}

\section{Oxygen vacancies}

In the following we want to investigate the effects of oxygen vacancies.
In order to allow for a more continuous variation of the concentrations,
also with access to low deficiency percentages that within the super
cell approach are beyond our computational power, we use the VCA.

It is not straight forward to simulate the absence of an atom within
the VCA. Ref.~\onlinecite{Shanavas2015} used a virtual mixture of
20\% of F with O to achieve an excess charge that turned out to electron-dope
the FeSe layer. However, while this replacement may simulate the effective
charge doping well, the local disordered potential of the interface
is likely different. Alternatively one could change the nuclear number
of e.g.~the Ti atoms to simulate dangling bonds due to the presence
of vacancies \cite{Richter2013}. Here, we choose to mix the oxygen
pseudo potential with vacuum, i.e. we scale the potential and the
charge with a parameter $x$.

All the atoms in the top STO layer are replaced with this virtual
potential. The evolution of the structure parameters of the interface
are shown in Fig.~\ref{fig:StructChargeNM}. We subtract from a value
shown in Fig.~\ref{fig:StructChargeNM} the value at zero percentage
such as $\Delta\alpha(x)=\alpha(x\%)-\alpha(0\%)$. As can be seen,
the variations of the FeSe internal structural parameters are almost
unaffected by the vacancy doping for low concentrations. As an interesting
effect, we find that instead of binding closer, at first for low vacancy
concentrations, the FeSe moves away from the substrate upon vacancy
doping. This trend is most pronounced in PBE, even though PBEsol also
shows this behavior. Then, at an already large concentration of $20\%$,
the trend is reversed and the binding distance shortens again. We
show the difference $\Delta d_{{\rm {\scriptscriptstyle Se}}-{\rm {\scriptscriptstyle O}}}=z_{{\rm {\scriptscriptstyle Se}}}-z_{{\rm {\scriptscriptstyle O}}}$
as a function of vacancy doping. $\Delta d_{{\rm {\scriptscriptstyle Se}}-{\rm {\scriptscriptstyle O}}}$
follows $\Delta h_{{\rm {\scriptscriptstyle Ti}}{\rm {\scriptscriptstyle Se}}}$
and thereby reflects the fact, that the layer shifts upwards while
the substrate remains intact. At about $30$\%, $\Delta d_{{\rm {\scriptscriptstyle Se}}-{\rm {\scriptscriptstyle O}}}$
starts increasing while $\Delta h_{{\rm {\scriptscriptstyle Ti}}{\rm {\scriptscriptstyle Se}}}$
decreases. This points out that the Ti binds to Se and not any more
to O. This may have to be attributed to our unrealistic description
of the potential in this region by the VCA.

Even though the charge transfer to Fe is rather small {[}compare Fig.~\ref{fig:StructChargeNM}
b){]}, in the general trend vacancies electron-dope the Fe. On the
other hand, according to the Bader analysis, the charge on the Se
close to the STO first increases before it decreases below the initial
value. Variations are one order of magnitude larger as compared to
the Fe atom. Charges on the vacuum side of the FeSe layer are highly
oscillatory which is because of the low density and low density variation
in the vacuum so that the Bader analysis is problematic.

In Fig.~\ref{fig:BandsNonMagn}, we show the evolution of the bands
at the $\Gamma$ and the M point for PBE and PBEsol as a function
of O vacancy concentration. We find that the hole pocket at $\Gamma$
with the majority character $d_{zy}$ is barely modified at first
and starts to shift downwards only for rather large vacancy concentrations.
It crosses the Fermi level away from the $\Gamma$ point in all cases
considered. Similarly at the M point, the bands are getting slowly
electron doped. The differences between PBE and PBEsol are minor. 

\begin{figure}
\begin{raggedleft}
\begin{minipage}[t]{1\columnwidth}%
\begin{center}
\includegraphics[width=0.7\columnwidth]{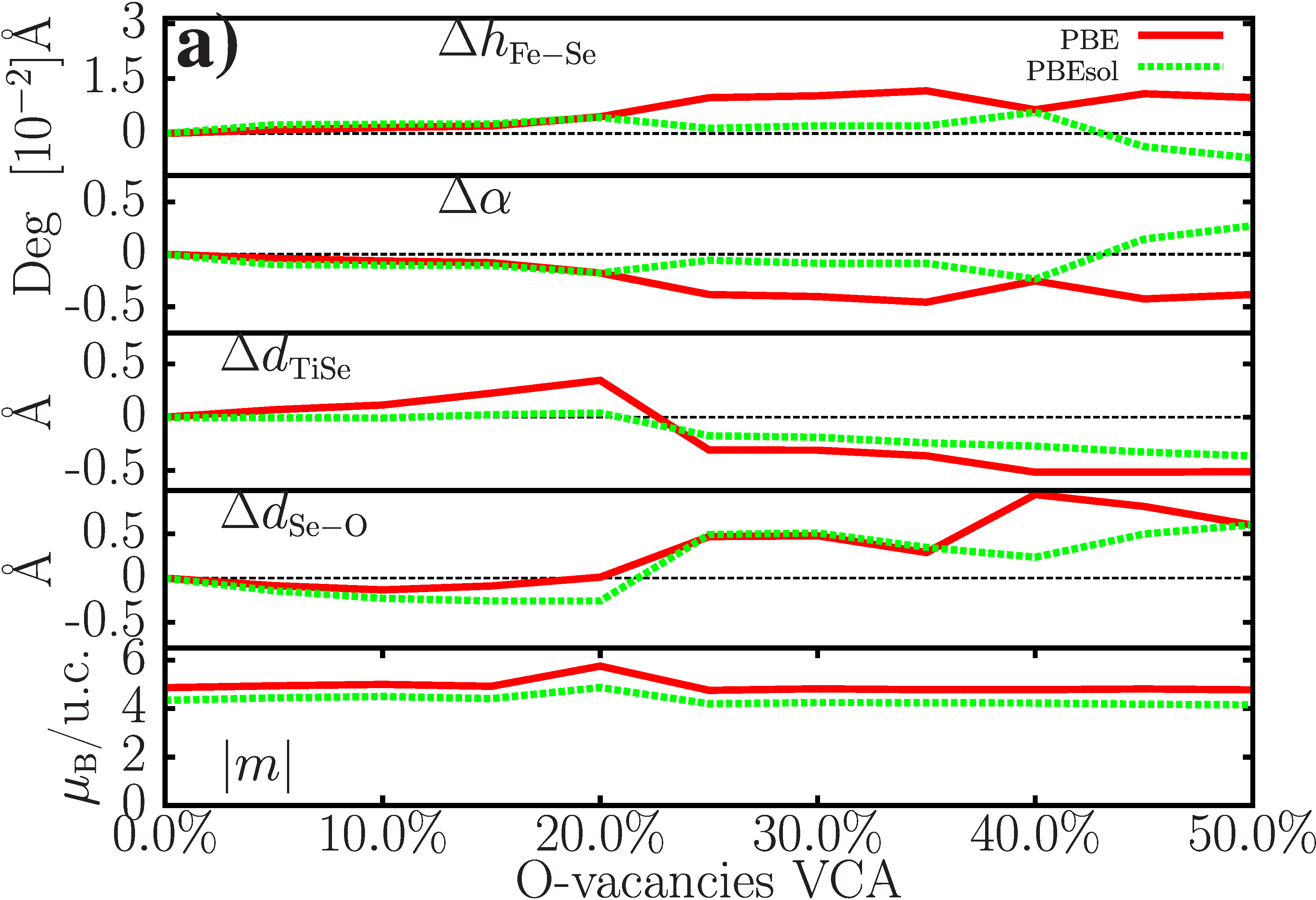}
\par\end{center}%
\end{minipage}\\
\begin{minipage}[t]{1\columnwidth}%
\begin{center}
\includegraphics[width=0.7\columnwidth]{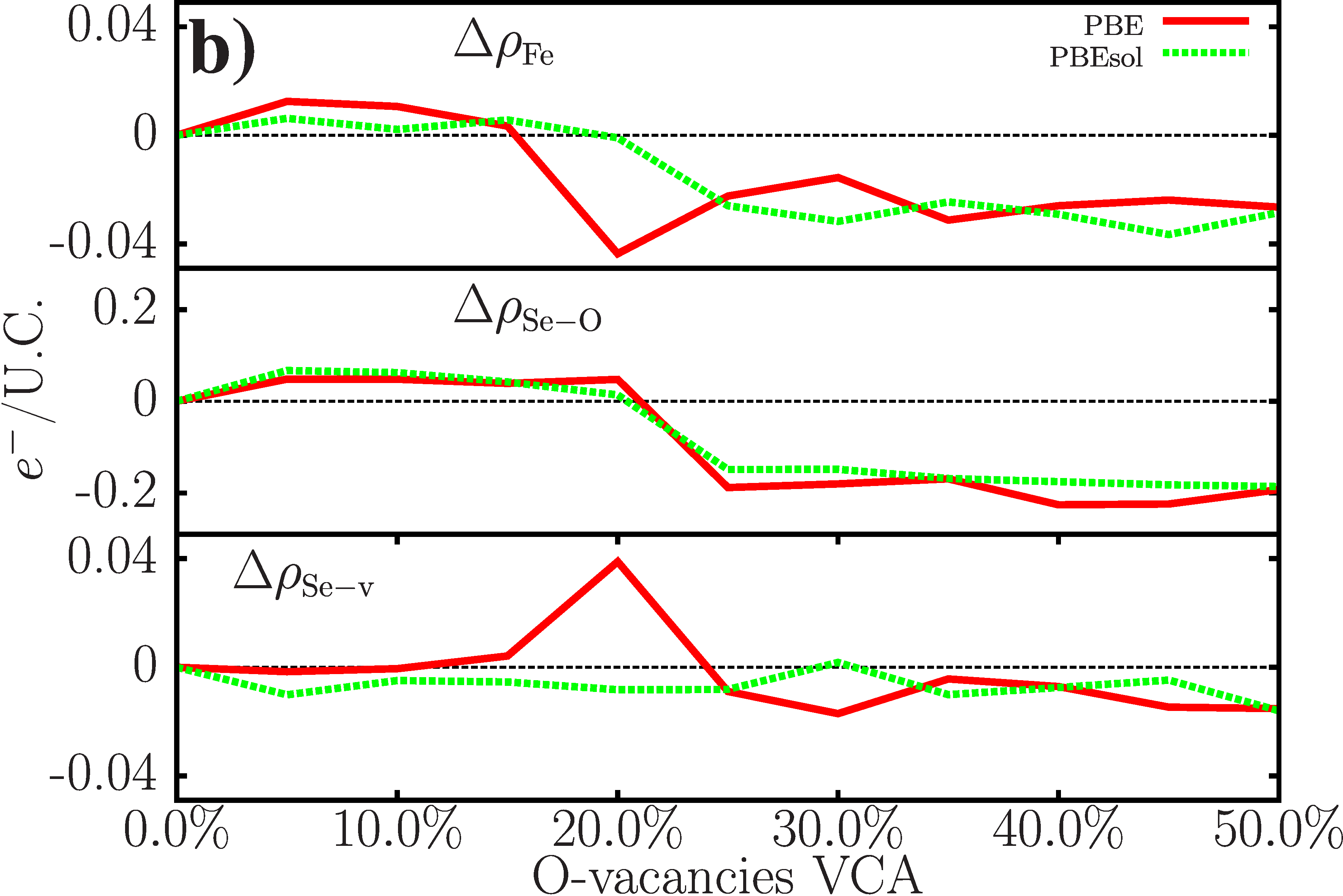}
\par\end{center}%
\end{minipage}
\par\end{raggedleft}

\caption{a) Structure data of the FeSe film within the AFM state as a function
of oxygen vacancies within the VCA. b) Charge polarization associated
to Fe and Se atoms according to a Bader analysis. Lattice parameters
are taken to be the experimental configuration.\label{fig:StructChargeM}}
\end{figure}

We repeat the analysis within the AFM state. As before, since the
numbers are close, we only perform the analysis for the experimental
lattice parameters. In Fig.~\ref{fig:StructChargeM},
we see that the structure follows a similar trend as the non-magnetic
state in the sense that vacancy doping has barely any effect on structure
parameters within the FeSe film. As before, we recover the behavior
of the increasing bonding distance of the FeSe film for low vacancy
concentrations. Again, the effect is only prominent in PBE.

As a side note, we observe that the magnetic moment is increased as
compared to a bulk calculation. Interestingly, there is a small antiferromagnetic
polarization induced in the first oxygen layer at 20\% vacancy doping.
The moments are of order 0.2$\mu_{{\rm {\scriptscriptstyle B}}}$/O
in PBE and thus contribute 10\% to the total magnetization. The moments
of the O are aligned with the Fe at the same $x-y$ position in the
unit cell. Otherwise the magnetization of the FeSe layer which itself
is only weakly affected by vancacy doping, does not extend to the
STO.

\begin{figure}
\begin{centering}
\begin{minipage}[t]{0.5\columnwidth}%
\begin{center}
\includegraphics[width=1\columnwidth]{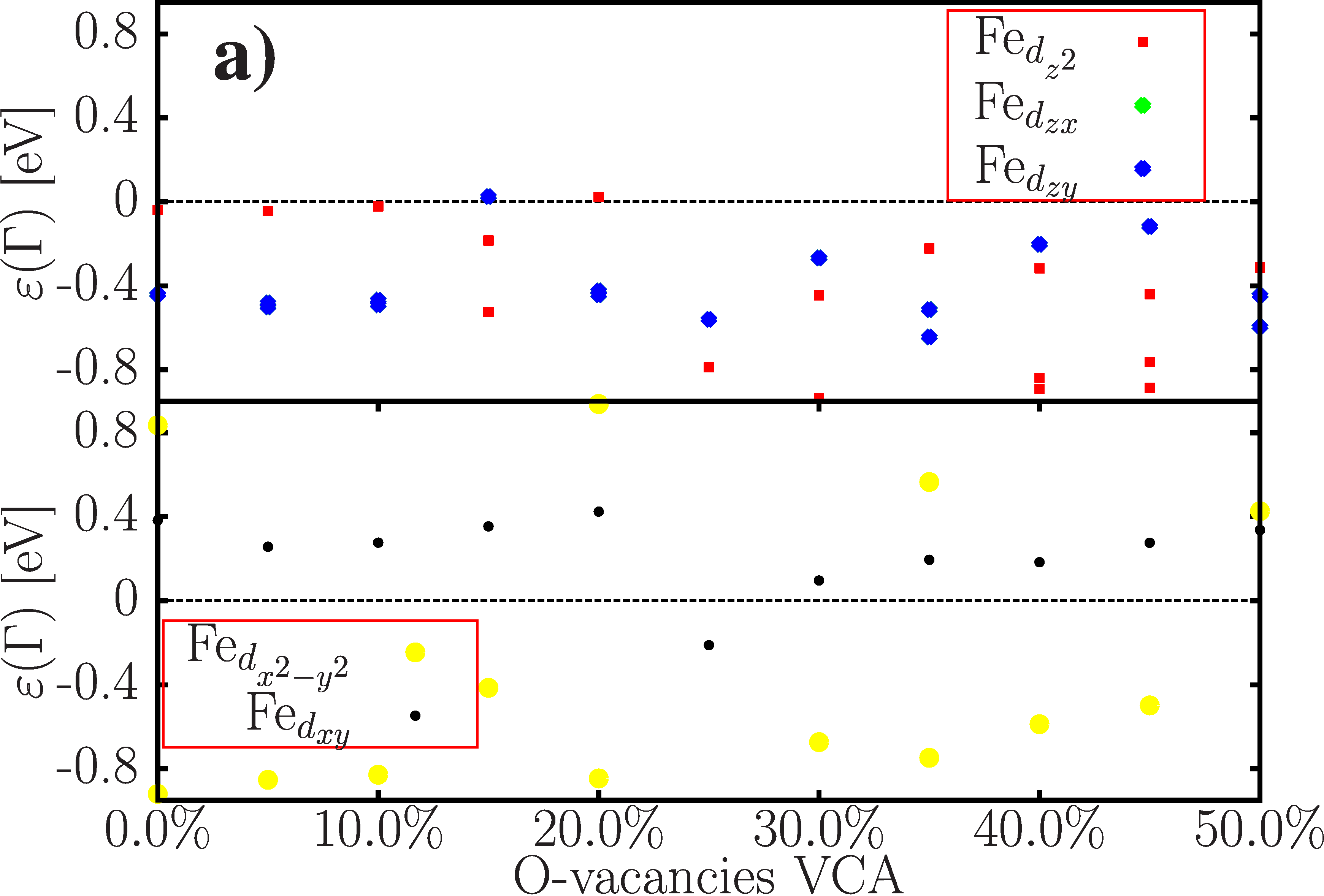}
\par\end{center}%
\end{minipage}\nolinebreak%
\begin{minipage}[t]{0.5\columnwidth}%
\begin{center}
\includegraphics[width=1\columnwidth]{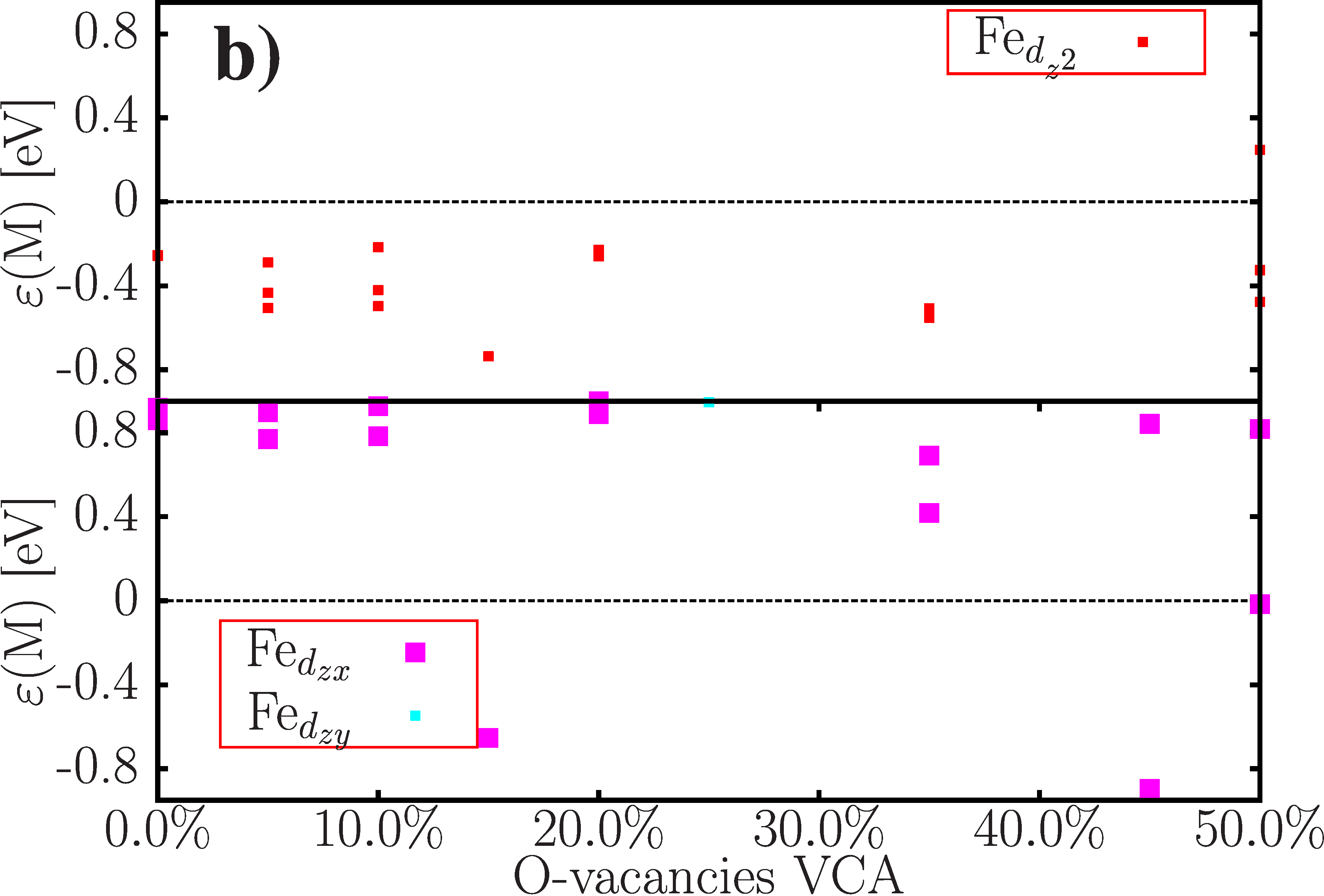}
\par\end{center}%
\end{minipage}\\
\begin{minipage}[t]{0.5\columnwidth}%
\begin{center}
\includegraphics[width=1\columnwidth]{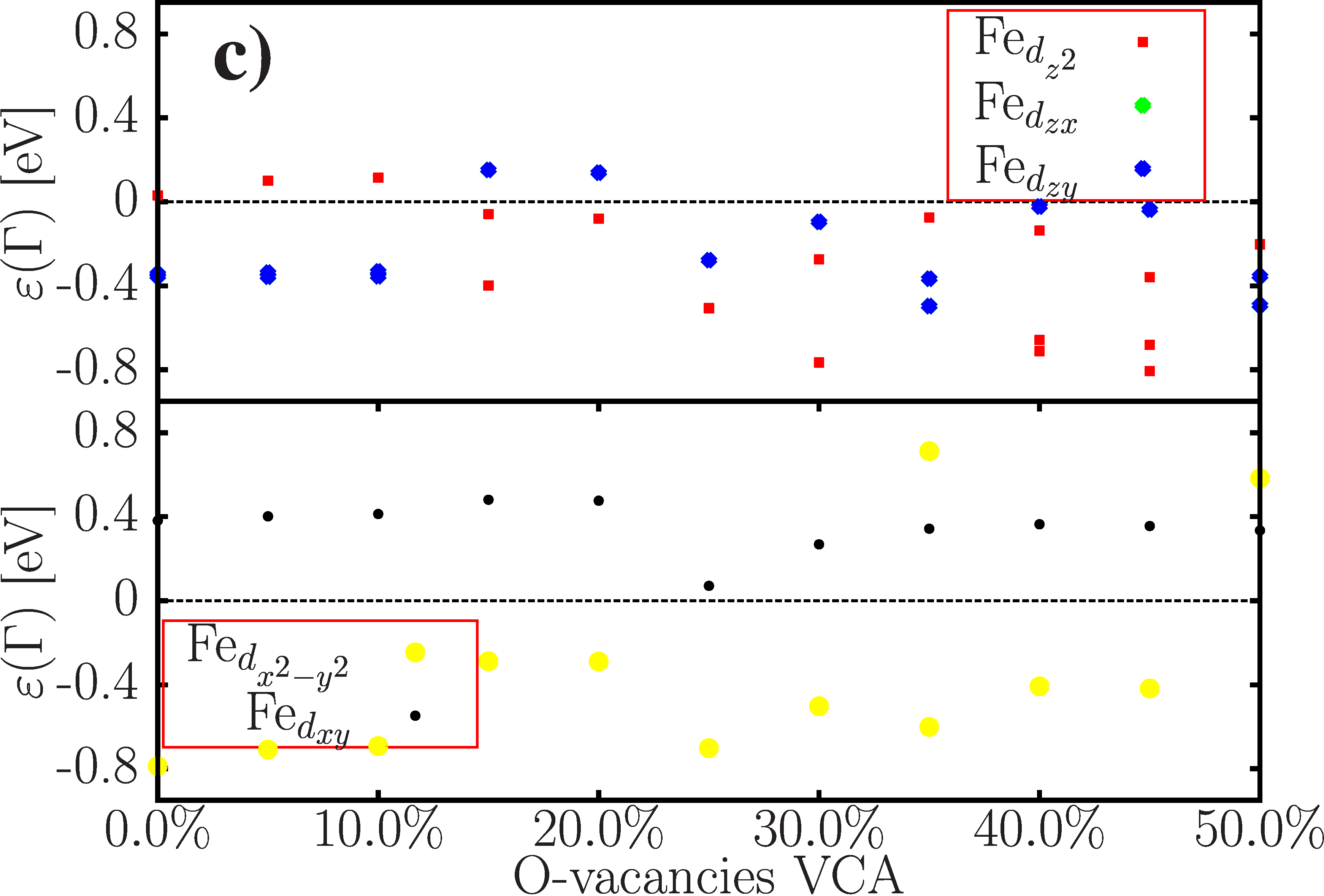}
\par\end{center}%
\end{minipage}\nolinebreak%
\begin{minipage}[t]{0.5\columnwidth}%
\begin{center}
\includegraphics[width=1\columnwidth]{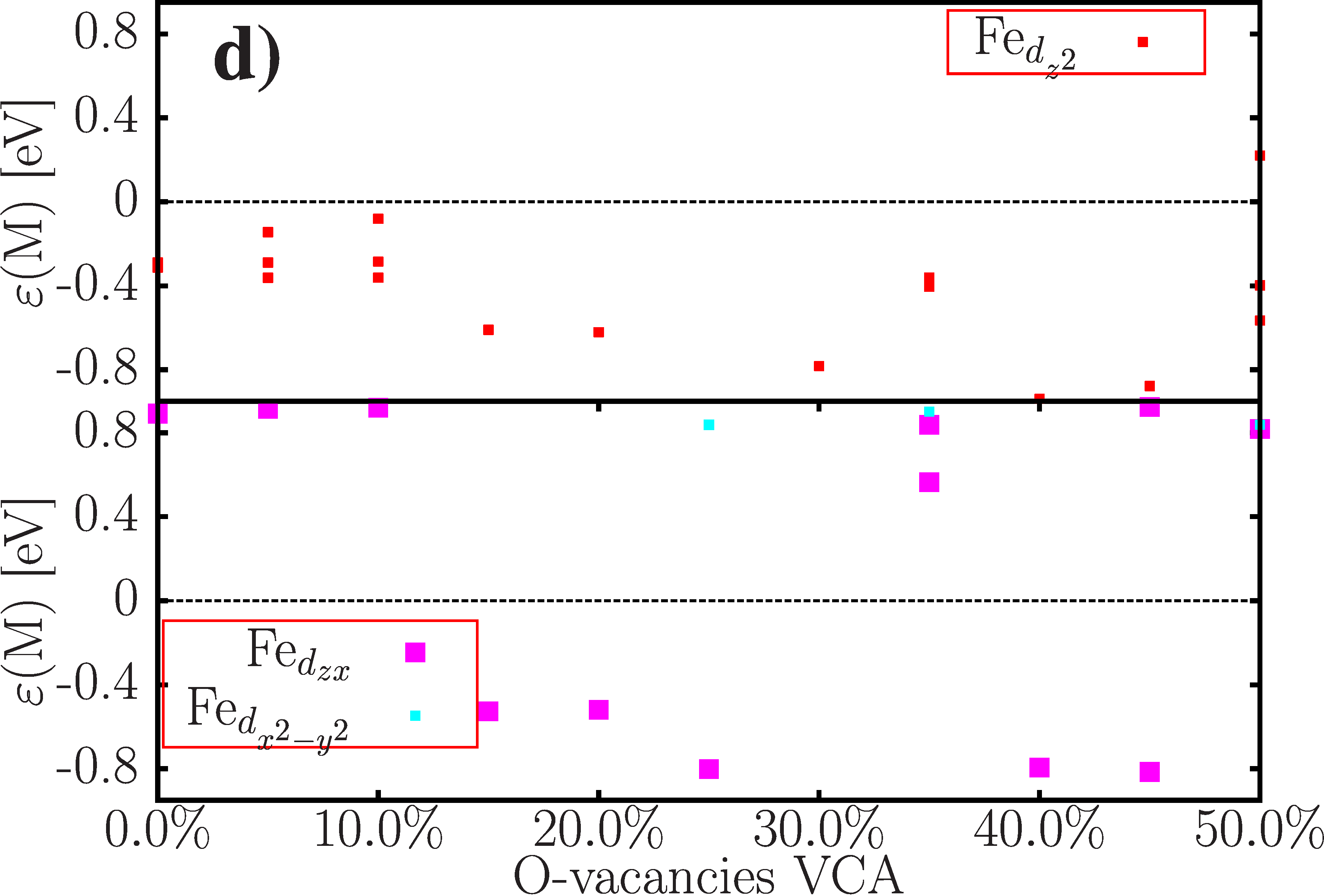}
\par\end{center}%
\end{minipage}
\par\end{centering}

\caption{Evolution of bands within the AFM state with a majority character
of Fe at $\Gamma$ {[}a) and c){]} and M {[}b) and d){]} point in
the Brillouin zone with O vacancy contribution. a) and b) are within
PBE while c) and d) are within PBEsol using experimental lattice parameters.\label{fig:BandsMagn}}
\end{figure}

The charges on the Fe site, on the other hand, first increase indicating
that the surface gets, in fact, slightly hole doped for very low concentrations
before the trend is turned around and the expected electron doping
is recovered. In Fig.~\ref{fig:BandsMagn}, we investigate the behavior
of the bands at $\Gamma$ and M as a function of vacancy concentration.
As was to be expected from the Bader analysis, there is only a minor
effect on the Fe $d$ orbital bands at the Fermi level for low vacancy
concentrations. Starting at about 25\%, the hole band at $\Gamma$
sinks below the Fermi level. 

\begin{figure}
\begin{centering}
\includegraphics[width=0.090909\columnwidth]{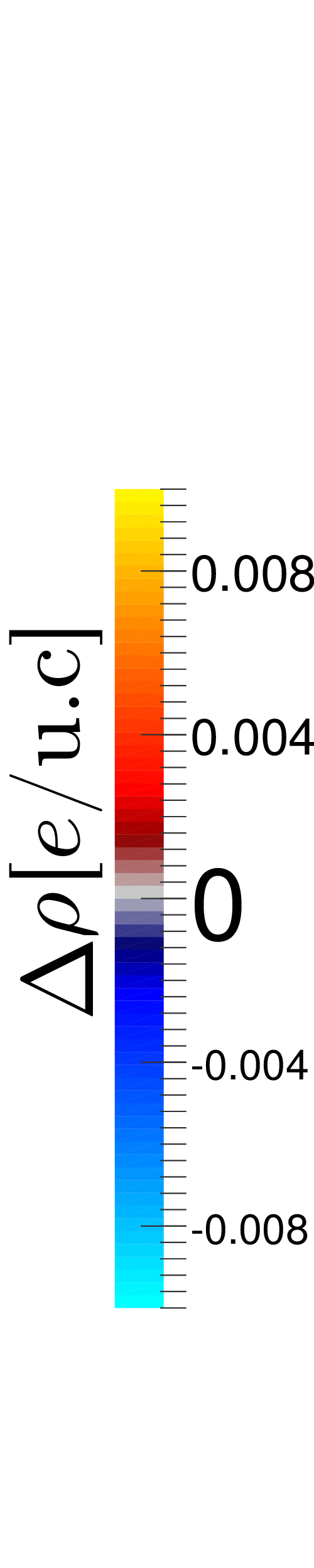}\includegraphics[width=0.181818\columnwidth]{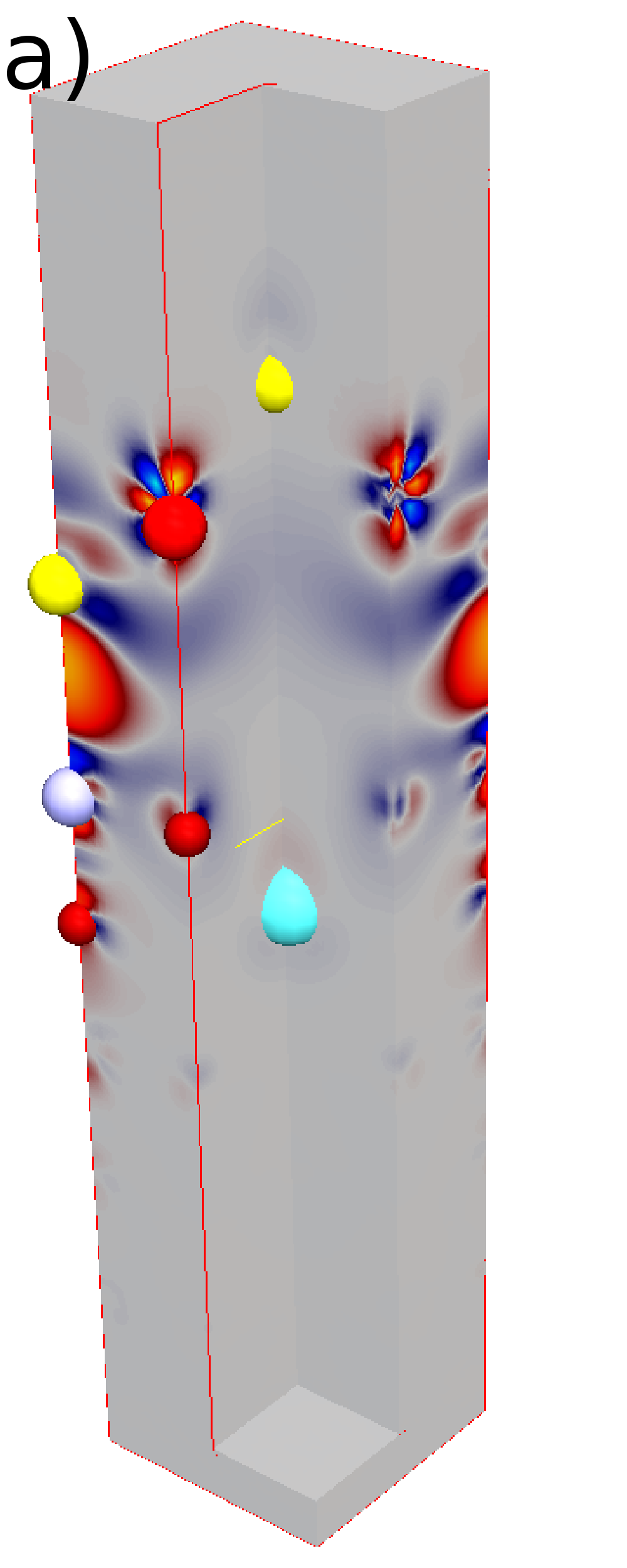}\includegraphics[width=0.181818\columnwidth]{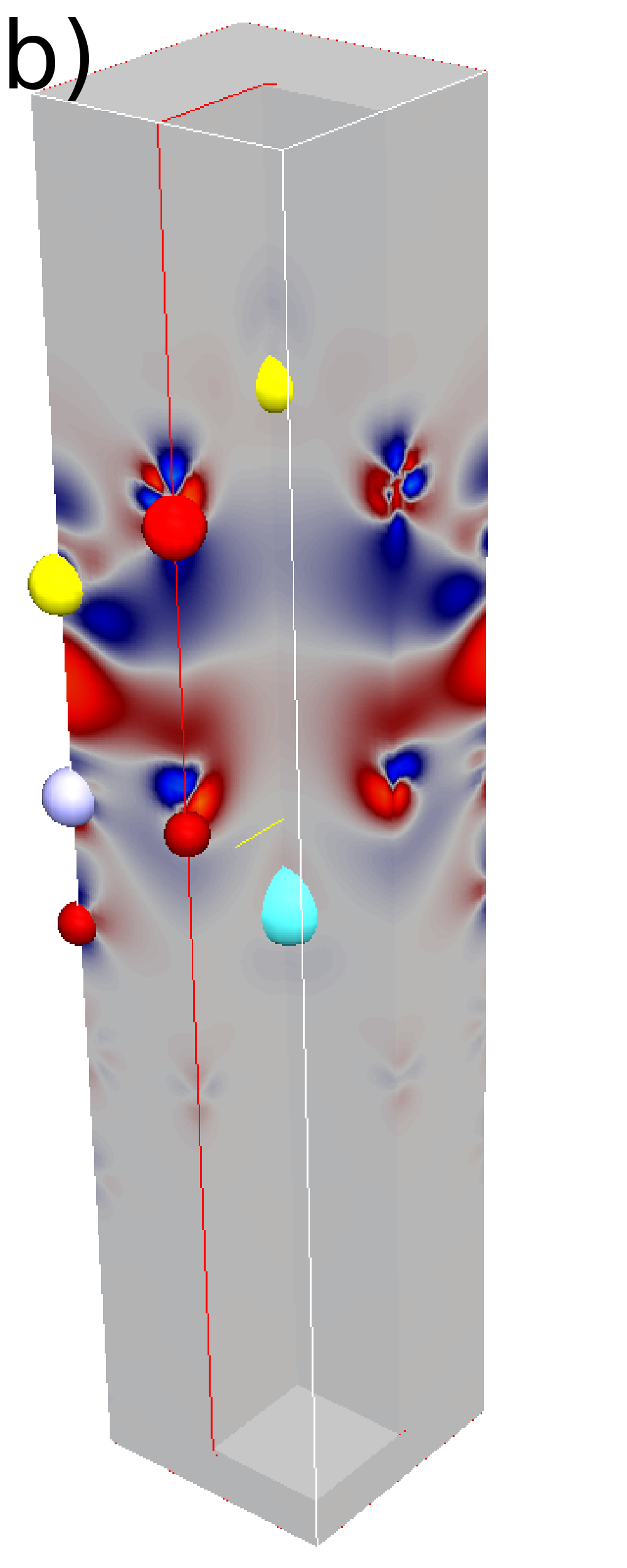}\includegraphics[width=0.181818\columnwidth]{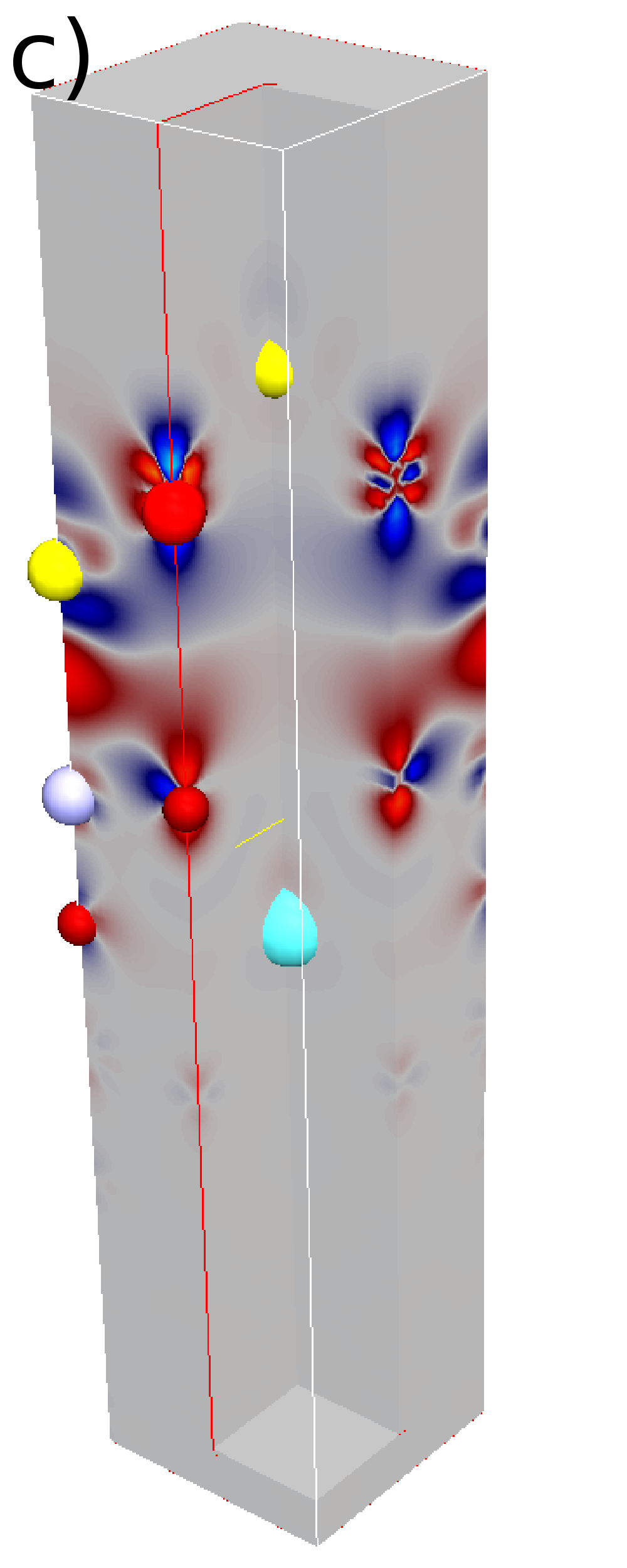}\includegraphics[width=0.181818\columnwidth]{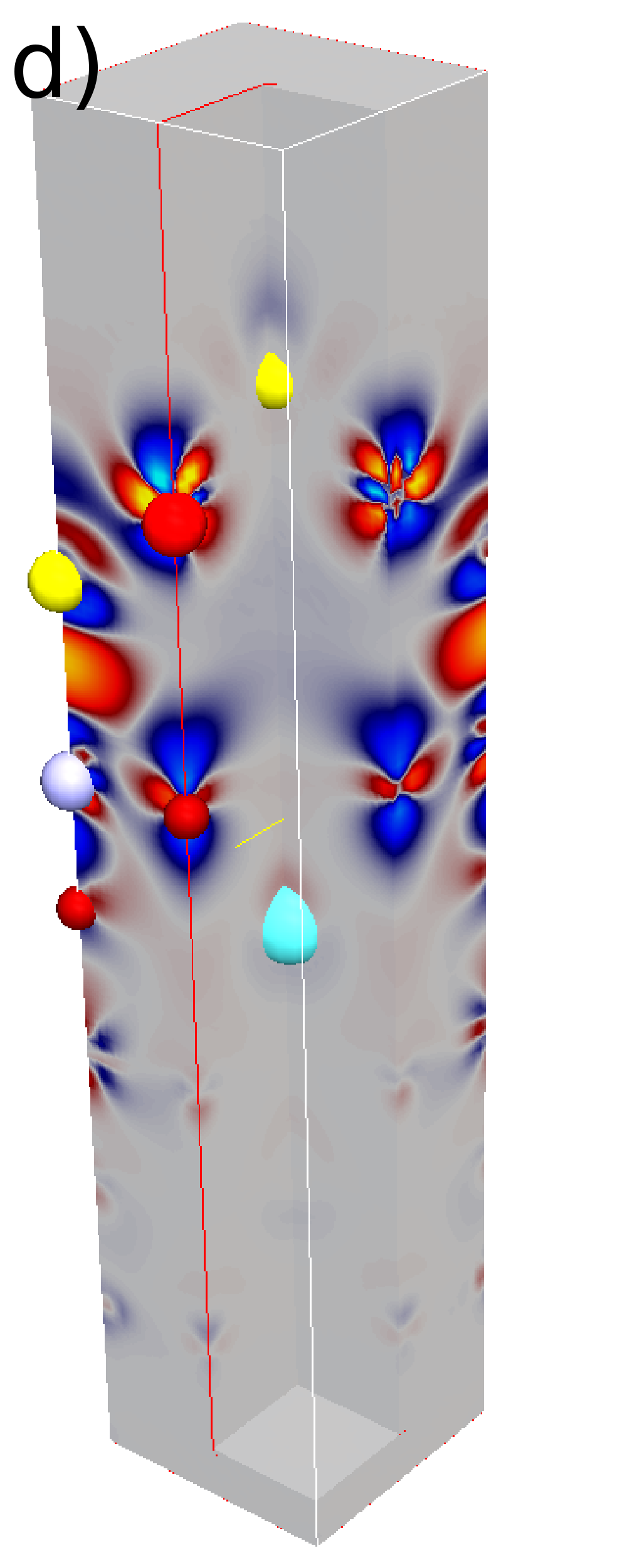}\includegraphics[width=0.181818\columnwidth]{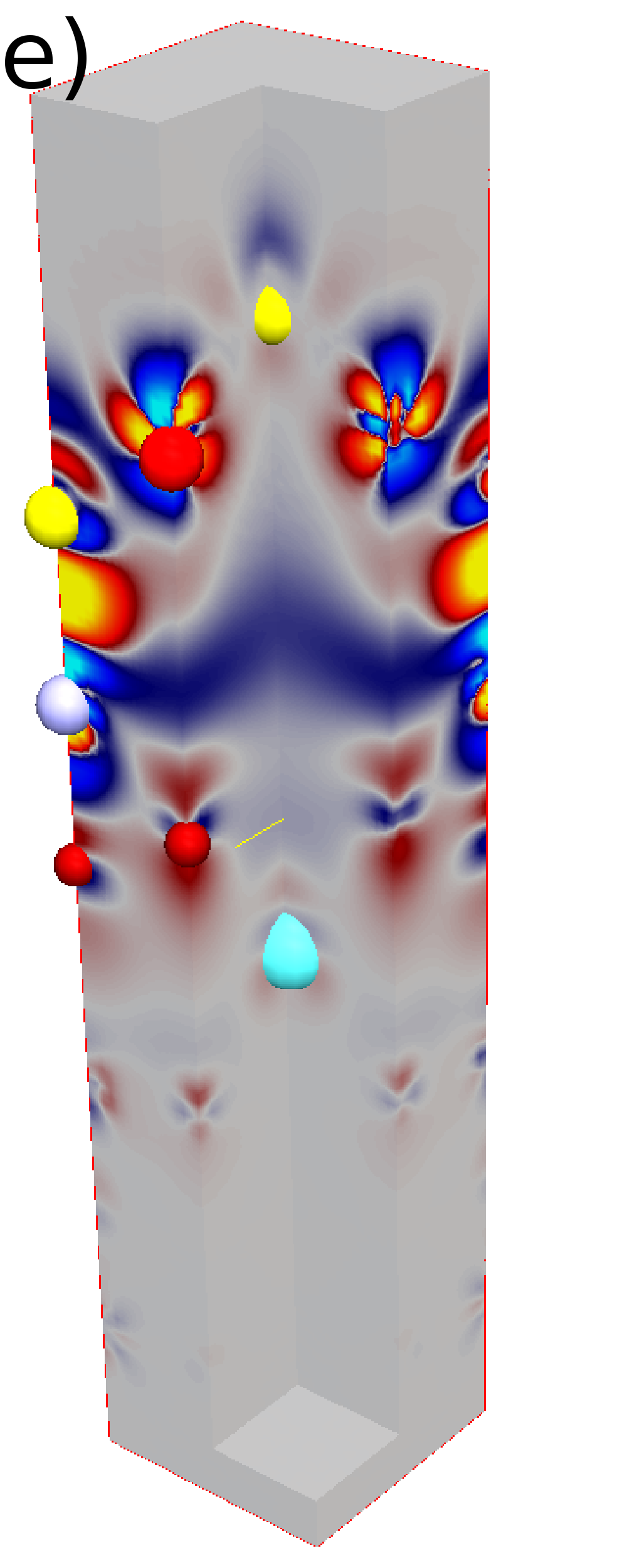}\\
\includegraphics[width=0.090909\columnwidth]{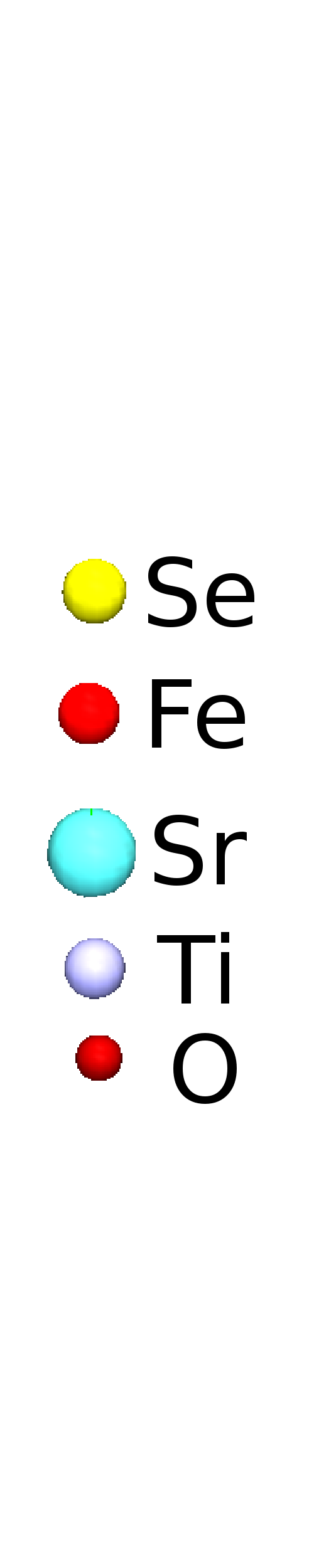}\includegraphics[width=0.181818\columnwidth]{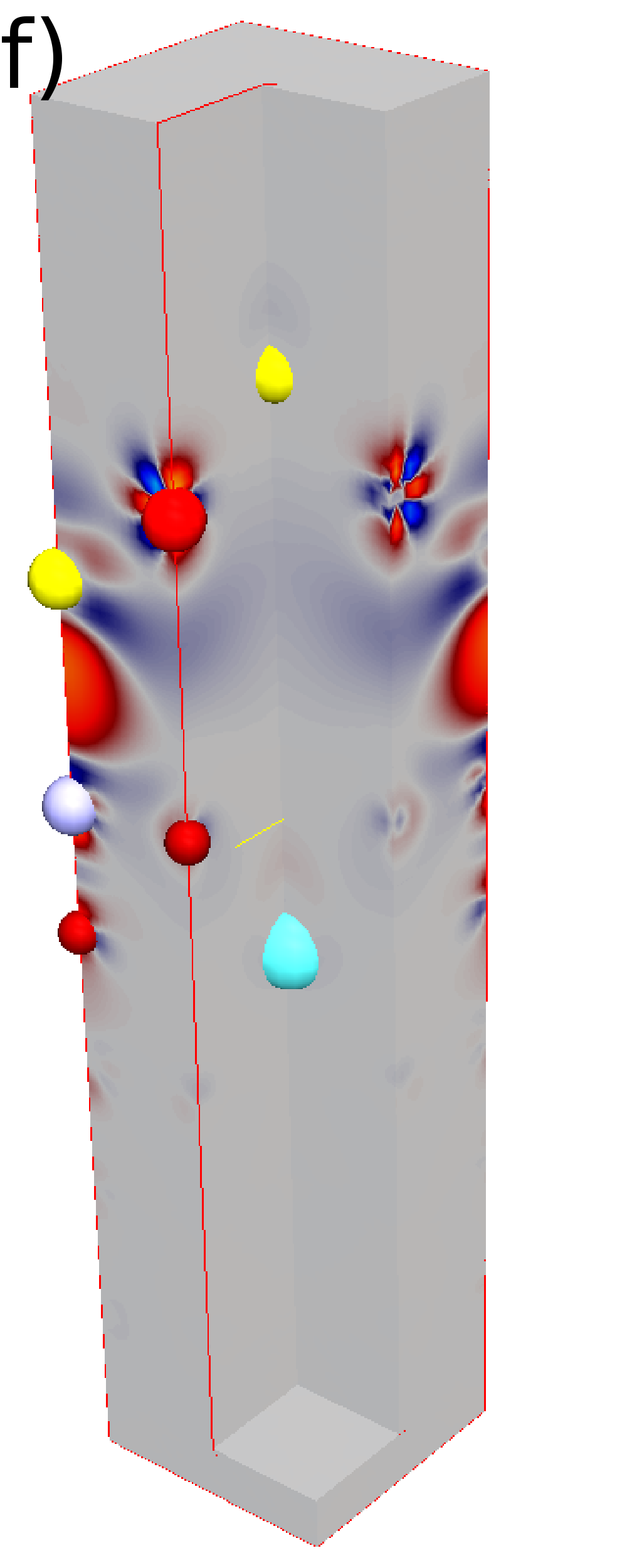}\includegraphics[width=0.181818\columnwidth]{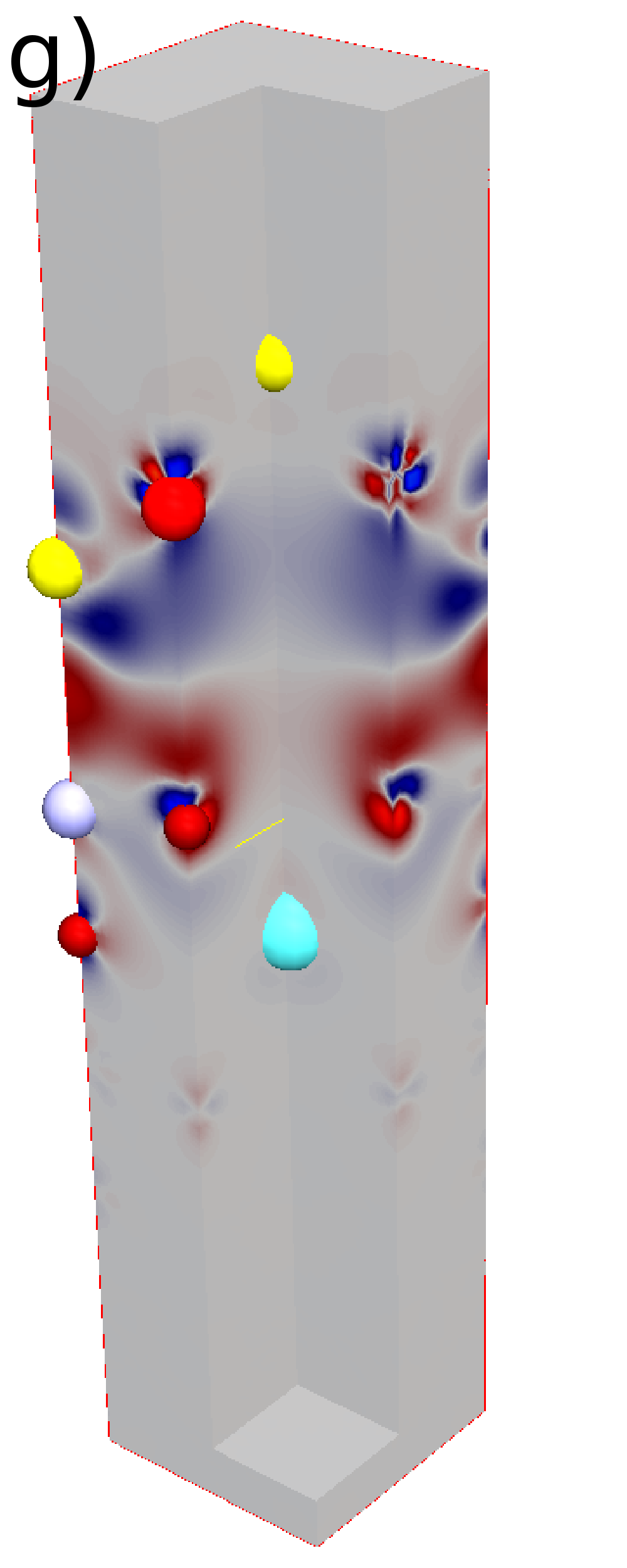}\includegraphics[width=0.181818\columnwidth]{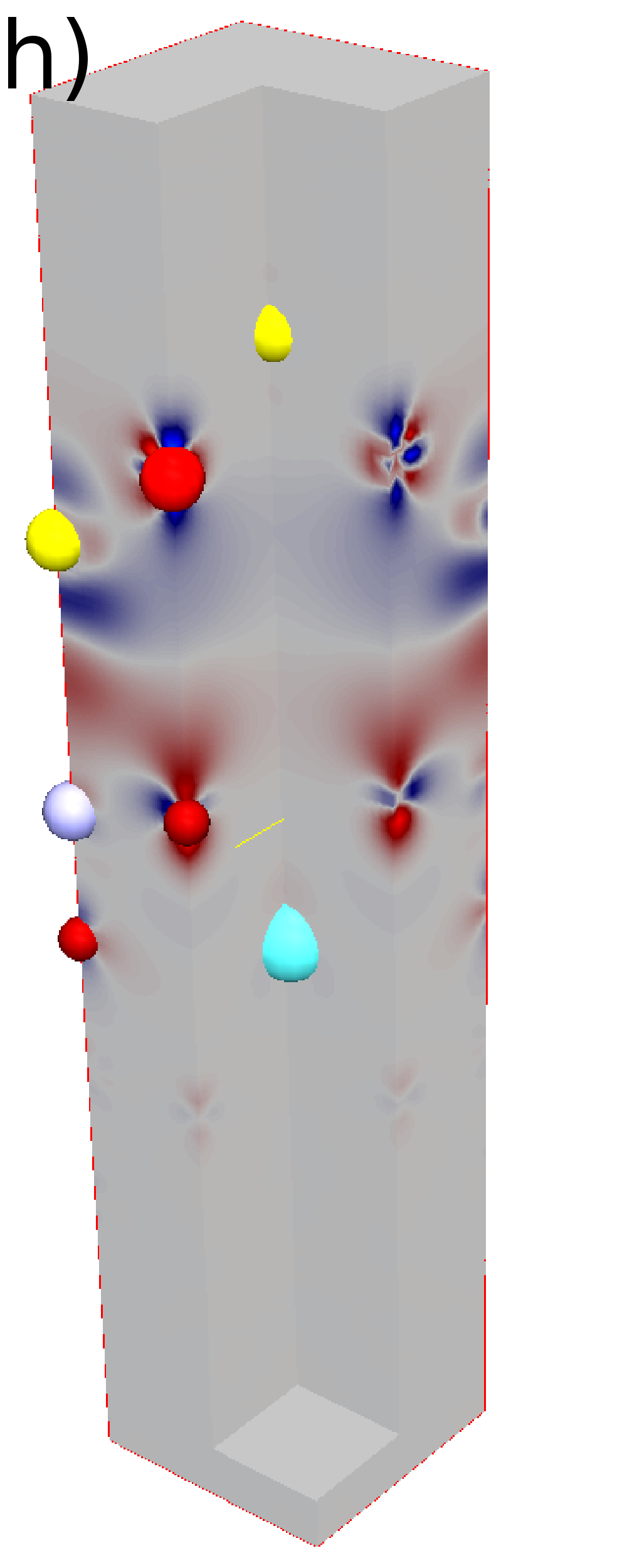}\includegraphics[width=0.181818\columnwidth]{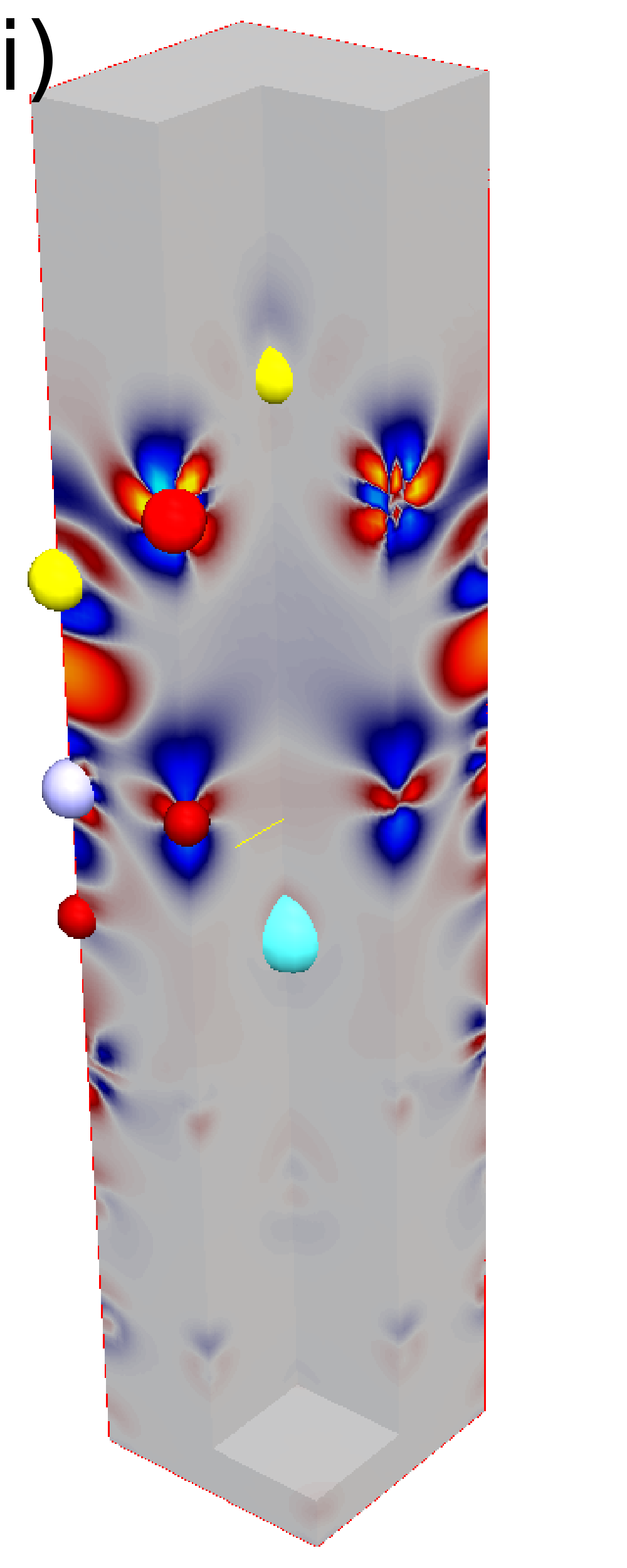}\includegraphics[width=0.181818\columnwidth]{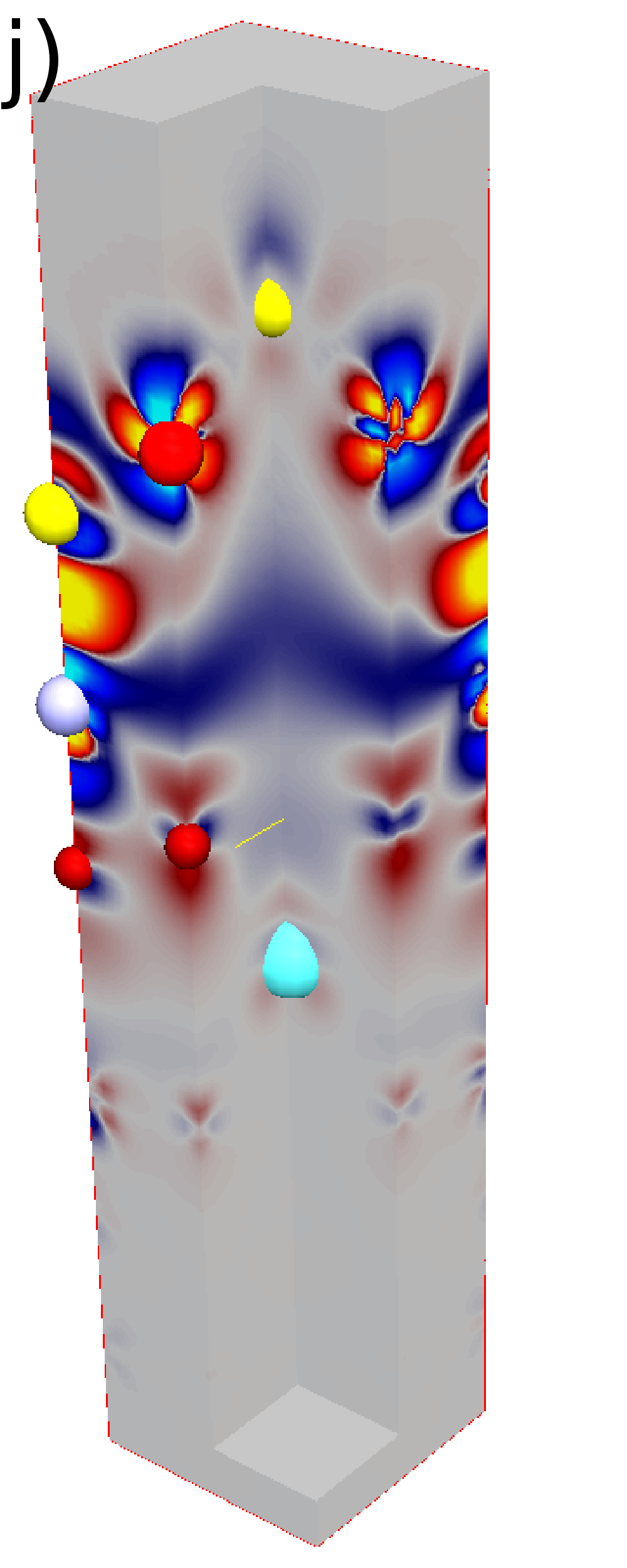}
\par\end{centering}

\caption{Development of the bonding structure with increasing oxygen vacancy
doping. We show the difference in densities between 1) the combination
surface and substrate minus 2) the sum of individual densities of
surface and substrate computed without the respective other one present.
Thus, electronic density is moved from blue to red regions due to
the interaction of surface and substrate. a-e) is computed within
PBEsol while f-j) is within PBE. a),b),c),d) and e) and f),g),h),i)
and j) correspond to the vacancy doping steps 0,10,20,30 and 40\%,
respectively. The position of some of the atoms in the unit cell are
shown as colored spheres in the plot.\label{fig:BondingStructure}}
\end{figure}
We conclude this analysis by highlighting the nature of the chemical
bonding structure that causes the initial unexpected increase of the
bond length with vacancy concentration. In Fig.~\ref{fig:BondingStructure},
we show the differences of first the combined layer plus substrate
where we subtract the sum of individual densities of surface and substrate
alone. This way, we can see where the chemical bonding of atoms redistributes
charge. As we focus on the interface interlayer bond length, we perform
this analysis in the NM state. We show the evolution of the bonding
structure in steps of 10\% oxygen vacancies, on the top in Fig.~\ref{fig:BondingStructure}
for PBEsol and on the bottom for PBE. The structure is, again, very
similar for PBE and PBEsol. For the stoichiometric interface {[}Fig.~\ref{fig:BondingStructure}
a) and f){]}, we find a predominately covalent bond, where the charge
is pulled in between the Ti and the Se atom. There is also a charge
redistribution among the Fe $d$ orbitals visible. Then, the bond
obtains an ionic character at $10\%$ and $20\%$ {[}Fig.~\ref{fig:BondingStructure}
b)-c) and g)-h){]}, where charge from the Se part of the FeSe layer
is pulled to the STO region before a more covalent nature reappears
in Fig.~\ref{fig:BondingStructure} d) and i) while for even higher
vacancy concentrations of 40\% {[}Fig.~\ref{fig:BondingStructure}
e) and j){]} charge is predominantly pulled from the STO region to
form a covalent bond of Ti with Se that also has a partly ionic charracter
visible by the blue regions close to the Ti atom. The only significant
difference between the two functionals is the strength of the bond,
which turns out to be weaker in PBE.

\section{Discussion and conclusion}

In conclusion, we find that the electronic and lattice structure of
the FeSe monolayer on STO in the antiferromagnetic state is in much
better agreement with experimental findings than the non magnetic
calculation. This goes hand in hand with the observation that the
binding distance is strongly underestimated in the non-magnetic state
and thus puts a unrealistically large artificial strain on the FeSe
in a calculation that uses bulk STO as a lattice constant. PBEsol
seems to be predicting the layer internal structure best. While magnetism
is important to describe the internal structure of the FeSe layer,
the binding distance to STO is weakly affected by it. Here, however,
we observe large differences in the functionals following their usual
hierarchy from short to long bond lengths: LDA, PBEsol and PBE. Compared
to recent experimental data\cite{Li2015}, and assuming the non-superconducting
sample is in the 0\% oxygen vacancy limit, the distances are all to
short with the minimal error of 6.3\% in PBE. Comparing the calculated
step height from the FeSe covered to uncovered STO areas with the
original experimental value of 5.5\r{A}\cite{Wang2012}, we find
that PBEsol agrees best. As an interesting observation we find that
the FeSe layer separation from STO in fact increases upon increasing
the oxygen vacancy concentration. The effect is present in PBE as
well as PBEsol even though the effect is weak in the latter case.
Assuming that the annealing steps in Ref.~\onlinecite{Li2015} are
mainly causing oxygen vacancies to form, this is in line with experiment
where the separation increases significantly from $\sim3.34$ to $\sim3.57$\r{A}.
We believe that this effect is caused by the tendency of O vacancies
to reduce the strength of the covalent bond Se to Ti at low concentrations.
In the calculations, PBE tends to soften the covalent bonds at a point
where the charge transfer due to vacancies have has not created a
compensating ionic bonding. Ordered vacancies in a super-cell calculation
may locally distort the bonding structure much more than the same
vacancy percentage in the disordered experimental system where we
would expect a more isotropic influence on electronic properties.
This may explain why a super cell calculation results in a strengthening
of a covalent bond that is suppressed if the deficiencies is disordered.
The homogeneous averaged potential, however, is better described within
the VCA.

Note also, that the incommensurate alignment of STO with the FeSe,
as seen in the data of Ref.~\onlinecite{Li2015}, may weaken the
covalent bonding of FeSe to STO and, thus, explain why PBE calculations
of the bond length agree better with experiment in our commensurate
approximation of the system.

We believe that more experimental data would be desirable to investigate
the precise nature of the interface. For example, Ref.~\onlinecite{Li2015}
concluded that the STO is terminated by a double layer of ${\rm Ti}-{\rm O}_{x}$.
Here, we found that a relaxation within DFT starting from the suggested
positions leads to an incorrect description of the system where the
double layer ${\rm Ti}-{\rm O}_{x}$ bind closely together leaving
the FeSe monolayer almost unbound.

Ref.~\onlinecite{Zhou2016} suggested that the slight Nb doping of
STO is important to understand the charge transfer and the electron
phonon coupling. In a future calculation it would be interesting to
investigate the effect of Nb impurities further. Especially in the
low vacancy region, our calculation finds that the FeSe bands are
almost unaffected. The presence of impurity states my well enhance
the electron transfer to STO and thus affect, both the resulting binding
distance as well as the charge transfer. On the other hand, Ref.~\onlinecite{Zhang2014}
showed that SC is similarly found in a FeSe monolayer on insulating
STO substrate, without Nb doping.

Furthermore it would be desirable to validate the findings of this
work with alternative methods to investigate the effect of disordered
oxygen vacancies such as the coherent potential approximation.
\begin{acknowledgments}
We have enjoyed stimulating discussion with S.~Johnston, P.J.~Hirschfeld,
Y.~Wang, T.~Berlijn, S.~Maiti and D.B.~Tanner. This work was supported
by Grant No.~DE- FG02-05ER46236.
\end{acknowledgments}
\bibliographystyle{apsrev4-1}
\bibliography{references}

\end{document}